# An *in situ* self-adaptive hydrogel coating enables seamless neural interfaces *via* okra mucilage polysaccharide and α-helical peptide amphiphiles co-assembly

**Authors**

Tenglong Luo[1,2#], Yiqing Guo[1,2#], Shanshan Su[1,2], Qiaoyu Yang[3,4], Wen Deng[5], Zhangfeng Huang[6], Zhiquan Yu[1,2], Dawen Yu[1,2], Yubin Ke[7], Hua Yang[7], Jiecong Wang[1,2*], Dewen Zhang[3,4,8*] and Yuanhao Wu[1,2*]

[1]Plastic and Reconstructive Surgery Department, Wuhan Union Hospital, Tongji Medical College, Huazhong University of Science and Technology, Wuhan, China

[2]Plastic and Reconstructive Surgery Research Institute, Wuhan Union Hospital, Tongji Medical College, Huazhong University of Science and Technology, Wuhan, China

[3]Department of Biophysics, School of Basic Medical Sciences, Health Science Center, Xi'an Jiaotong University, Xi'an, China

[4]Institute of Medical Engineering, Translational Medicine Institute, Xi'an Jiaotong University, Xi'an, China

[5]Department of Neurology, Tongji Hospital, Tongji Medical College, Huazhong University of Science and Technology, Wuhan, China

[6]Thoracic Surgery Department, Wuhan Union Hospital, Tongji Medical College, Huazhong University of Science and Technology, Wuhan, China

[7]Spallation Neutron Source Science Center, Dalang, Dongguan, China

[8]School of Future Technology, Xi'an Jiaotong University, Xi'an, China

#These authors contribute equally.

*Corresponding authors:

Jiecong Wang wangjiecong1982@sina.com, Dewen Zhang zhangdewen@xjtu.edu.cn, Yuanhao Wu yuanhaowu@hust.edu.cn.

**Abstract**

Long-term stability of neural interfaces is frequently compromised by mechanical mismatch and chronic neuroinflammation, often leading to electrode detachment and signal failure. While hydrogel coatings offer a solution, conventional designs typically rely on exogenous conductive fillers that can sacrifice mechanical flexibility or induce toxicity. Here, we report on a soft neural interface based on the supramolecular co-assembly of a renewable natural polysaccharide, okra mucilage polysaccharide (OMP), and an α-helical peptide amphiphiles (APA). The resulting OMP-APA hydrogel (OP gel) exhibits environment-responsive enhancements in bioadhesion and charge-transport capability triggered by physiological pH and electrical stimulation. These properties arise from intrinsic, stimulus-responsive alterations in fibre architecture and orientation, eliminating the need for conductive fillers. Leveraging interfacial liquid-liquid phase separation, we demonstrate the *in situ* coating of ultra-thin (2.8 ± 0.3 μm) OP-gel coating onto carbon fibre electrodes (CFE).

The OP-gel-coated electrodes (OP-CFE) significantly mitigate foreign body responses and glial scarring, enabling stable, high-quality neural recordings in a mouse cortical *in vivo* model. Our findings provide a versatile strategy for constructing seamless, multifunctional bio-interfaces through supramolecular co-assembly, with broad implications for advancing neural prosthetics and neuroscience research.

**Introduction**

Neural electrodes serve as critical interfaces that connect the brain to external electronic devices[1, 2], enabling the treatment of neurological disorders[3], brain-machine interface research[4], and fundamental neuroscience exploration[5, 6] through electrical stimulation or recording neural activity. However, the long-term stability and performance of electrodes are often undermined by chronic degradation arising from material-tissue mismatch, including mechanical property[7, 8], limited material stability[9], poor biocompatibility[10]. In addition to impairing recording and stimulation performance, a more critical concern is that long-term incongruence leads to electrode detachment including electrode displacement[9, 11] and glial scarring[12]. These issues can cause interface failure[13], often necessitating revision surgery and resulting in increasing the risk of procedure-related trauma, recovery time, and economic burdens for patients[14-16]. Consequently, developing materials that match the mechanical and biological properties and stability of brain tissue is essential for ensuring the long-term performance and stability of neural electrodes, aiming to enhance electrode functionality and reduce the risk of detachment.

A key strategy to ensure long-term electrode stability is to strengthen the adhesion integrity between the electrode and the host neural tissue[17] through adding coating layers. Such reinforcement thereby diminishes the likelihood of detachment triggered by physiological micromotion[9] and adverse tissue responses[18]. However, coating materials also need mechanical modulus, biocompatibility, and immunomodulatory function[19, 20], which can decrease the risks of foreign body response and chronic neural inflammation[10]. In this context, hydrogel-based neural electrodes have emerged as a promising strategy to address these multifaceted challenges[21-23]. Hydrogels typically exhibit excellent biocompatibility, bioadhesive properties[24], mechanical compatibility[25], and dynamic morphological adaptability[26]. These properties can promote seamless integration, ultimately forming a stable electrode-tissue interface[27]. But a key limitation is that conventional hydrogels typically possess low electrical conductivity[28], which requires strategies such as doping of conductive nanocomposites[29] to meet the operational demands of neural interfacing applications. This approach often alters the mechanical properties of the hydrogel, such as reducing its flexibility and elasticity[30], and may induce cytotoxicity[31], oxidative reaction[32], or release harmful substances[33, 34], potentially triggering local toxicity and inflammatory responses[35]. Therefore, an ideal solution necessitates the development of a hydrogel that intrinsically combines bioadhesiveness with high charge-transport capability, without relying on the addition of exogenous conductive fillers.

Supramolecular co-assembly enables monomers to enhance structural complexity[36, 37] while retaining their individual characteristics[38] *via* non-covalent interactions[39], which presents a promising avenue for developing advanced hydrogel coatings for neural electrodes. For instance, Nam *et al.* co-assembled a betaVhex peptide with carbon nanotubes into a hydrogel neural interface[40], while Jain Deepak *et al.* wrapped gold electrode using a glycosylated nucleoside fluorocarbon-based supramolecular gel[41]. However, these strategies still need conductive fillers. Peptide amphiphiles (PAs) are particularly compelling building blocks in this context. They have been shown to confer conductive properties by direct self-assembly behavior[19, 42] or sequestering conducting polymers during the assembly process[43, 44]. Furthermore, we have previously designed co-assembling hydrogels based on PAs and polysaccharide[36], protein[45], or human-derived complex components[46], which proves that PAs can be a powerful tool to manipulate co-assembly with a diversity of matters. Additionally, these co-assembled materials are compatible with skin[47], vascular[37] or adipose tissues[48], demonstrating the feasibility of extending co-assembly strategy to develop materials for softer neural tissue.

Based on the requirements of neural tissue applications, the development of supramolecular co-assembled neural electrodes involved the use of our research group's independently designed α-helical peptide amphiphiles (APA), which have been validated to carry charges and are capable of constructing fibre modules with hierarchical structures[48]. In addition, to construct a multifunctional hydrogel that addresses the key requirements of neural interfaces, we put our finger on a renewable natural candidate: okra mucilage polysaccharide (OMP). Derived from okra (Abelmoschus esculentus), with a long history of traditional medicinal use[49], OMP has been identified as a primary bioactive macromolecule[50]. It exhibits superior biocompatibility[51], anti-inflammatory[52] and antioxidant activities[53] both *in vitro* and *in vivo*. Additionally, OMP possesses excellent adhesive performance[54] and shear-thinning behavior[55] in aqueous solutions. Its identified chain-like polymer structure[50] suggests a high potential to serve as a co-assembly templated monomer with PAs[36]. Therefore, OMP and APA offer a potential to fabricate a supramolecular hydrogel coating layer, designed to enhance neural electrode performance.

Here, we report on an innovative soft neural electrode based on natural OMP and APA. The resulting OMP-APA hydrogel (OP gel) exhibits environment-responsive enhancements in adhesion and charge-transport capability, triggered by physiological cues of *in situ* pH and electrical stimulation. This functionality stems from intrinsic, stimulus-responsive alterations in the co-assembled fibre architecture[56, 57] and orientation[58], rather than incorporation of extra adhesive moieties or conductive dopants, thereby preserving the inherent biocompatibility and mechanical properties of the material. Also, the OP gel exhibits significant anti-inflammatory properties derived from OMP. Employing interfacial liquid-liquid phase separation (LLPS) strategy, we uniformly coat an ultra-thin OP hydrogel layer down to $2.8 \pm 0.3$ μm *in situ* onto carbon fibre electrodes (CFE) by a facile dip-coating process[59]. Subsequently, in a mouse cortical electrode implantation *in vivo* model, the OP-gel-coated electrodes (OP-CFE) has been demonstrated to effectively mitigate foreign body response, local neuroinflammation and prevent glial scarring, while achieving stable and high-quality neural signal recording. Taken together, we have pioneered the use of supramolecular co-assembly technology to construct an *in situ* interfacial layer to improve the neural electrodes. This soft neural electrode achieves seamless structural and functional integration between the implant and neural tissue, and shows wide potential as a new material for advancing neuroscience research and neural interface technology.

**Results**

## 1. Rationale of the system and materials

A co-assembled hydrogel system was designed based on natural plant polysaccharides and PAs to create coating layers for neural electrodes. OMP was selected as the polysaccharide component due to its high propensity for chain entanglement[50] and anti-inflammatory property[52]. To further enhance the system's structural programmability and responsiveness, we synthesized a custom peptide amphiphile APA consisting of a hydrophilic peptide sequence (AEKIRKE) conjugated to a hydrophobic tail (Fig. 2a)[48]. The α-helical structure of APA peptide sequence enables the formation of close-packed, axially aligned electronic states through co-assembly[60, 61]. Coupled with the supramolecular order of the system, this arrangement collectively endows the co-assembled material with long-range charge-transport capability[62, 63]. Through liquid-liquid phase separation (LLPS)-driven interfacial co-assembly, OMP and APA form a nanofibrous OP-gel that undergoes structural transformations triggered by pH changes and electrical stimulation, which correlate with the macroscopic adhesion[56, 64] and charge-transport properties[58] of the coating. By leveraging the LLPS-driven interfacial diffusion-reaction process, we fabricated nanofibrous OP-gel coating directly on neural electrode surfaces (Fig. 1a, flowchart), leading to rapid *in situ* enhancement of charge-transfer ability (Fig. 1a, the region highlighted by the dashed box). The resulting coating exhibits adaptive electrode-tissue adhesion (Fig. 1b) and enhanced charge-transfer efficiency (Fig. 1a, bottom) upon implantation-relevant pH shifts and electrical stimulation. Concurrently, the coating effectively mitigates local neuroinflammation induced by mechanical mismatch and foreign-body implantation (Fig. 1c). Together, these adaptive functional improvements foster seamless integration at the electrode-brain tissue interface and enable high-fidelity, stable neural signal recording.

## 2. Characterization of the OMP-APA co-assembly system

*LLPS-driven interfacial hierarchical co-assembling process*

To validate the potential for LLPS between the two components, zeta potential was used and revealed that OMP and APA exhibited the opposite zeta potential (-12.6 mV vs. 26.7 mV) (Fig. 2a). In addition, gel permeation chromatography (GPC) revealed that OMP has a peak molecular weight of $4.78 \times 10^3$ kDa, which is orders of magnitude larger than that of APA (1.11 kDa). These pronounced differences in surface charge and molecular size provide the molecular basis for triggering and sustaining LLPS-mediated supramolecular co-assembly[65]. Then, the liquid-liquid interface co-assembly was driven by directly injecting an APA aqueous solution into an OMP saline solution (Supplementary Movie 1), which was triggered immediately upon contact and led to the formation of a robust soft three-dimensional OP gel (Fig. 2d). Furthermore, to elucidate the structural origin of the mechanical integrity, scanning electron microscope (SEM) observations revealed continuous fibrous network of the surface of OP gel (Fig. 2e, left). The cross-sectional views, which is parallel to the diffusion direction, showed an orderly stacked lamellar structure (Fig. 2e, right), indicating that the supramolecular network develops in a highly directional manner because of the interfacial diffusion-reaction mechanism[37, 66]. To further visualize the spatial distribution of molecular co-assembly during the LLPS process, OMP was labeled with red fluorescent particles and APA was conjugated with FITC. The strongly overlapping signals in fluorescence images reveal that the two components form a continuous and uniform fibrous network parallel to the interface (Fig. 2f), which is consistent with the SEM images. Collectively, these results demonstrate that OMP and APA undergo LLPS-driven interfacial co-assembly, yielding a robust nanofibrous hydrogel with a well-defined lamellar hierarchy.

*Nanoscale characterization of OMP-APA co-assembling materials*

To elucidate the interplay between OMP and APA during co-assembly, we systematically compared the nanoscale structures of two components and their co-assembled hierarchical architectures. Firstly, the conformations of this system were confirmed by transmission electron microscopy (TEM). OMP exhibits a disordered, chain-like morphology, forming highly entangled filaments of varying diameters under the experimental conditions, with the finest fibres measuring 4.2 ± 1.0 nm (Fig. 2g, left). In contrast, APA self-assembles into uniform and well-defined fibres, indicative of a highly ordered supramolecular assembly (Fig. 2g, middle). Upon co-assembly, OMP and APA organize into high-aspect-ratio ribbon-like nanostructures, which exhibit longitudinally aligned alternating bright and dark striations. Notably, twisting was observed in the nanoribbons (Fig. 2g, right, yellow arrows), demonstrating their ability to maintain structural integrity under deformation. And tapping-mode atomic force microscopy (AFM) provided consistent topographical information (Fig. 2h). Interestingly, AFM imaging resolved that APA self-assembles into uniform helical fibres displaying clear M-twisted morphology (Fig. 2h, middle and Supplementary Fig. 2). This chiral supramolecular twist likely originates from the transfer of the intrinsic helical chirality of the APA peptide at the molecular level to the supramolecular assembly[67]. Furthermore, the co-assembled nanostructure features of OMP-APA were revealed by small-angle neutron scattering (SANS) (Fig. 2i and Supplementary Fig. 3). Fitting with a flexible-cylinder model showed that OMP alone exhibited cylindrical nanofibrils with 151 Å radius, while APA self-assembles into flexible chains with 59 Å radius. After co-assembling, the scattering curve was well described by a flexible-cylinder model with 38 Å radius, suggesting the formation of reorganized elementary fibrillar units. At the low-$q$ region, the scattering curves exhibited a clear $q^{-1.6}$ dependence, indicative of further aggregation and interconnection of the elementary fibrillar units into an open, branched, and tenuous 3D network[68]. These nanoscale findings are consistent with the hierarchical fibrous morphology observed by TEM and AFM. In conclusion, these results demonstrate that OMP and APA cooperatively co-assemble into high-aspect-ratio nanofibrillar building blocks and further extend into an interconnected three-dimensional network.

*Mechanical properties of OP gel after co-assembling*

The observed structural integrity and softness of the OP gel prompted further investigation into its mechanical properties. In subsequent experiments, since OMP alone lacks hydrogel-forming capability, a hydrogel formed by the self-assembly of APA was prepared and used as the control group. Frequency sweep measurements (Fig. 2j) confirmed solid-like elasticity in both hydrogels (G' > G'' across 0.1-100 rad s$^{-1}$). The OP gel exhibited a lower storage modulus (G') than the APA gel (6.2 kPa vs. 8.8 kPa), better matching the mechanical modulus of brain tissue (0.1-1.5 kPa)[69], especially compared with the carbon fibres or other traditional materials of neural electrodes (~GPa)[11, 70] (Fig. 1c, top). Concurrently, the higher loss modulus (G'') of OP gel (4.3 kPa vs. 2.4 kPa) indicated its macroscale mechanical toughness and biomimetic energy-dissipating capabilities, which shows particular promise for applications in flexible bioelectronics. In strain-recovery tests, the OP gel recovered 48.6% of its original G' within 1 minute, markedly exceeding the 4.0% recovery of the APA gel (Fig. 2k), demonstrating rapid self-healing capability. This dynamic reversibility stems from the abundant non-covalent interactions within the gel[71], allowing for rapid reorganization in physiological environments and ensuring long-term stability. In conclusion, these results confirmed that the supramolecular co-assembling of OMP with APA yields a hydrogel whose mechanical modulus closely matches that of brain tissue, while also exhibiting excellent energy-dissipation and rapid self-healing capabilities. These properties endow the OP gel with the potential to mitigate foreign-body responses and to accommodate the dynamic physiological environment of brain tissue.

### 3. *In situ* pH responsive adaption of the OMP-APA co-assembly system

*pH-responsive properties of OMP and APA*

Given the ionisable groups (e.g., -COOH from Gal-UA and Glc-UA) in OMP (Supplementary Fig. 1, bottom) and the amphoteric nature of APA, we investigated the pH sensitivity of electrostatic interactions that govern their self-assembly across physiologically relevant pH conditions. To characterize this responsiveness under conditions relevant to implantation, transitioning from an *in vitro* state (pH ≈ 6.0) to the mildly alkaline brain extracellular environment (pH ≈ 7.3)[72] (Fig. 1b), we analyzed the self-assembly behavior of OMP and APA by dynamic light scattering (DLS) at pH 4, 6, and 8. DLS analysis revealed pronounced pH-dependent aggregation of OMP in aqueous solution, with the average particle size increasing from 1.2 ± 0.3 μm (pH 4) to 2.8 ± 1.6 μm (pH 6), and further to 5.1 ± 1.6 μm at pH 8 (Fig. 2b). Similarly, APA exhibited pH-responsive self-assembly, with its average particle size increasing from 0.33 ± 0.04 μm (pH 4) to 0.45 ± 0.07 μm (pH 6), and 0.83 ± 0.05 μm at pH 8 (Fig. 2c). These results demonstrate that both OMP and APA undergo systematic, pH-dependent self-assembly, providing a supramolecular basis for the enhanced fibre entanglement and elongation in the co-assembled system under the elevated pH conditions mimicking the brain physiological environments.

*pH-responsive micro- and nano-structural reorganization of OP gel*

Given that supramolecular co-assembly relies on non-covalent interactions[36, 37], the intrinsic pH-responsiveness of OMP and APA is expected to be retained in the co-assembled system [73]. We therefore examined whether the OP gel undergoes structural reorganization at both the micro- and nano-scale across the physiologically relevant pH range. Firstly, SEM revealed a persistent, interconnected fibrous network across the tested pH range (pH 4/6/8; Supplementary Fig. 4a). Quantitative analysis showed that the fibre diameters shifted from 27 nm (pH 6) to ~20 nm (pH 4/8) (Supplementary Fig. 4b), indicating that the network remains continuous while undergoing pH-driven microstructural modulation. Then TEM and AFM were employed to achieve higher resolution of these transitions. At the synthesis pH of 6, the OP gel exhibited nanoribbon-like fibres with diameters of 22-25 nm (Fig. 3c, d, middle). Upon increasing the pH to 8, the morphology transitioned into a more densely packed network characterized by entangled helical fibres with reduced diameters (13-16 nm; Fig. 3c, d, right). Significantly, TEM characterization at pH 7.3, mimicking the brain extracellular environment, showed a fibre architecture indistinguishable from that at pH 8 (Supplementary Fig. 5). This structural convergence justifies the use of pH 8 as a representative experimental condition for the deprotonated physiological state. In contrast, acidification to pH 4 induced a transition from elongated fibres into fragmented, short-fibre clusters (Fig. 3c, d, left). SANS further demonstrated pronounced pH-dependent nanoscale reorganization (Fig. 3b). Consistent with our earlier findings, the scattering profile at pH 6 reflected a hierarchical network of fibres with a 38 Å radius. At pH 8, low-$q$ scattering retained a $q^{-1.6}$ dependence, indicative of an open, branched 3D network, while high-$q$ flexible-cylinder fitting revealed a reduced fibril radius (~27 Å), reflecting subtle thinning of individual fibres. In contrast, at pH 4, low-$q$ scattering exhibited a $q^{-3}$ dependence, and high-$q$ Guinier-Porod fitting revealed a substantially larger radius (~424 Å), indicating the formation of large, loosely packed, disordered clusters rather than individual fibrils. Collectively, these multi-scale observations confirm that the OP gel undergoes pronounced, pH-driven reorganization of its co-assembled fibrous network (Fig. 3a), and notably, under simulated brain microenvironment, the system forms a more entangled helical fibre network.

*pH-responsive bio-adhesion of OP gel*

Considering the importance of the morphology and packing density of fibrous networks in governing wet biological adhesion[56, 64], we postulated that the pH-dependent nanostructural reorganization of the OP gel would translate into tunable adhesive strength at the material-tissue interface. To validate this, adhesion forces of OP-gel films at pH 4, 6, and 8 were measured in liquid using AFM-based force spectroscopy (Supplementary Fig. 6). Force curves were acquired with a 20 μm-diameter colloidal probe and analyzed using the Johnson-Kendall-Roberts (JKR) model[74]. Peak retraction forces were used to quantify adhesion. At pH 6, the OP gel exhibited an average peak adhesion force of 1,400 pN. Acidic conditions (pH 4) reduced adhesion to 590 pN, whereas alkaline conditions (pH 8) increased it markedly to 4,738 pN (Fig. 4g). This trend was consist of the transition from fragmented or loosely associated fibrils to densely packed and entangled fibrous bundles observed at higher pH. These results confirmed that the OP gel can dynamically adapt its adhesive properties in response to pH changes, enabling it to enhance interfacial integration precisely within physiological environments.

Furthermore, we evaluated OP gel adhesion to a range of wet biological tissues, including brain, heart, liver, spleen, lung, and kidney, all of which readily adhered to the gel (Fig. 3f, Supplementary Fig. 7). To extend this assessment from the tissue level to the cellular scale under controlled conditions, Dil-labeled BV-2 cells, widely distributed brain-resident cells, were cultured on OP-gel films deposited on mica substrates and prepared under different pH conditions (pH=4/6/8). After 6-hour incubation, confocal fluorescence imaging revealed that cell adhesion was significantly enhanced on all OP-gel films compared with bare mica (291 cells mm$^{-2}$). In agreement with the AFM-based adhesion measurements, cell densities were 525 cells mm$^{-2}$ at pH 4, 601 cells mm$^{-2}$ at pH 6, and increased substantially to 1,079 cells mm$^{-2}$ at pH 8 (Fig. 3e, j). Collectively, these results demonstrate that the OP gel undergoes pH-dependent supramolecular reorganization that translates into enhanced wet-tissue and cellular adhesion upon implantation, with potential to maintain stable electrode-tissue integration *in vivo*.

## 4. Biological validation

*Biocompatibility of OP gel*

Biocompatibility is critical for neural implant coatings to minimize neuroinflammation and adverse tissue responses. Therefore, we first assessed the in vitro cytotoxicity and proliferative influence of the OP gel by co-culturing it with human umbilical vein endothelial cells (HUVECs) and BV-2 cells for 7 days, alongside APA-gel and untreated controls. These two cell types were selected to represent vascular and immune components that are critically involved in the tissue response to neural implants. Continuous proliferation was observed in both cell types across all groups (Fig. 4a, b). Live/Dead assays conducted in a 3D co-culture format further confirmed high viability, with HUVECs and BV-2 cells retaining typical morphologies and showing minimal cell death throughout the culture period (Supplementary Fig. 8 and Fig. 4c). Notably, the OP gel maintained a stable growth surface by day 7, whereas the APA gel exhibited curling and partial layer detachment (Fig. 4c), highlighting superior structural integrity of OP gel. We next assessed *in vivo* biocompatibility by subcutaneous injection of OP and APA gels in BALB/c mice for two weeks. Immunohistochemical analysis revealed uniform host-cell infiltration in both hydrogels, with the OP gel exhibiting more pronounced vascularization (Fig. 4j, k). Hematoxylin and eosin staining of major organs (heart, liver, kidney, spleen, brain) showed no apparent tissue toxicity for either gel (Supplementary Fig. 9). In conclusion, these results demonstrate that OP gel exhibits excellent *in vitro* and *in vivo* biocompatibility, robust structural stability, and favorable interfacial integration, supporting its suitability as a coating material for neural electrodes.

*3D growth of the HUVECs and BV-2 within the OP gel*

To elucidate how the co-assembled OP gel modulates cell-material interactions at the interface, we examined the three-dimensional morphology and cytoskeletal organization of HUVECs and BV-2 cells cultured within the OP gel matrix. The cell-material interactions were observed by SEM firstly. On day 1 and day 7, both cell types were able to infiltrate and grow within the fibrous network of the OP gel, adopting a three-dimensional (3D) growth mode rather than forming a planar monolayer (Fig. 4d). By day 7, HUVECs exhibited rounded cell bodies distributed throughout the gel matrix and formed frequent intercellular contacts, indicative of adaptation to the three-dimensional microenvironment. In contrast, BV-2 cells maintained a predominantly spherical morphology over the culture period, consistent with their reported behavior in soft, brain-mimetic matrices. Higher-magnification SEM images revealed extensive multi-point contacts between the co-assembled fibres and the cell membranes (Fig. 4d, 5k×), indicating that the OP gel provides abundant anchoring sites for cells and enables stable 3D adhesion.

To further characterize cytoskeletal organization, F-actin staining was performed on day 1, 3, and 7 (Fig. 4e). At day 1, both HUVECs and BV-2 cells displayed rounded morphologies with limited cytoskeletal extension, consistent with early-stage adaptation to the 3D environment. By day 3, partial cytoskeletal reorganization was observed, with HUVECs showing localized elongation and BV-2 cells beginning to extend cellular processes. By day 7, HUVECs maintained a stable three-dimensional morphology within the gel, while BV-2 cells developed an increasingly interconnected cytoskeletal network, indicative of enhanced cell-cell communication within the matrix. Collectively, these results demonstrate that the OP gel supports stable 3D cell adhesion and promotes cytoskeletal adaptation within the co-assembled fibrous network. Such biomimetic cell-material interactions are expected to facilitate intimate integration at the implant-tissue interface, which is critical for the long-term performance of hydrogel-coated neural electrodes.

*Anti-inflammation effect of OP gel in vitro*

Implantation of neural electrodes inevitably triggers microglial activation and local inflammatory responses[75, 76], which can compromise long-term device performance and lead to the final detachment. Given the intrinsic anti-inflammatory activity of OMP, we reasoned that this bioactivity would be retained in the co-assembled OP gel. To evaluate whether the OP gel could modulate microglial inflammatory activation, we established a classical *in vitro* neuroinflammation model of BV-2 microglial cells stimulated by lipopolysaccharide (LPS). Meanwhile, an OP-gel-loaded Transwell insert was introduced as an intervention group to partially recapitulate implantation-relevant inflammatory conditions (Fig. 4f). After 24 h of stimulation, nitric oxide (NO) release and intracellular reactive oxygen species (ROS) levels were quantified as representative markers of microglial inflammatory activation. LPS treatment markedly increased NO production compared with the untreated control, whereas co-culture with the OP gel significantly attenuated this LPS-induced NO overproduction (Fig. 4g). Consistently, LPS stimulation resulted in a pronounced elevation of intracellular ROS levels, while cells treated with Rosup exhibited the highest oxidative stress as a positive control. Notably, co-culture with the OP gel effectively suppressed the LPS-induced ROS increase, yielding significantly lower ROS levels than those observed in the LPS-only group (Fig. 4h). Importantly, this attenuation of inflammatory markers was not attributable to compromised cell viability, since BV-2 cells maintained normal proliferation and viability when co-cultured with OP gel as demonstrated above. Collectively, these findings demonstrate that the OP gel significantly mitigates LPS-induced microglial inflammatory activation *in vitro*, supporting its potential as an immunomodulatory coating for neural implants to reduce implantation-associated inflammatory responses.

## 5. Fabrication and interfacial assembly *in situ* of OP-CFE

*Visualization of the LLPS-driven coating process*

To elucidate the dynamic mechanism underlying the *in situ* formation of OP-gel coating, we visualized the co-assembly process in real time using confocal fluorescence microscopy with fluorescently labeled OMP and APA. Upon contact between OMP and APA droplets, rapid co-assembly was initiated at the liquid-liquid interface, forming a nascent interfacial layer with a thickness of $2.8 \pm 0.3$ μm (Fig. 5a, yellow arrows). Subsequently, APA molecules diffused across the initial interface and continuously interacted with unreacted OMP chains, driving spatially propagating gelation and 3D matrix growth (Fig. 5a, white arrows). This direct visual evidence confirms that OP-gel formation proceeds *via* an LLPS-mediated diffusion-reaction process, enabling rapid, conformal hydrogel coating directly on the electrode surface.

*In situ fabrication and morphological characterization of the OP-gel coating*

To translate this LLPS-driven co-assembly into a functional neural interface, we fabricated OP-CFE *in situ* using an innovative facile dip-coating process. Specifically, CFE tips were sequentially immersed in 1% APA aqueous solution and 0.5% OMP saline solution to build the coating through interfacial co-assembly (Fig. 5d). The resulting microstructure was characterized by SEM. While bare CFE tips exhibited a smooth surface with uniformly distributed fibre bundles (Fig. 5d, left and Supplementary Fig. 10a), OP-CFE tips displayed a conformal, continuous coating composed of densely packed co-assembled nanofibres, yielding a distinctly roughened surface morphology (Fig. 5d, right and Supplementary Fig. 10b). Quantitative thickness analysis performed on representative SEM images using ImageJ revealed that the OP gel formed an ultrathin coating layer with an average thickness of $3.0 \pm 0.1$ μm. To further validate the successful coating and its bioactivity, carbon fibres partially coated with OP gel were co-cultured with BV-2 cells for 6 h. Optical microscopy combined with live/dead staining revealed that cells preferentially adhered to the OP-gel-coated segments, whereas virtually no attachment occurred on bare carbon-fibre surfaces (Fig. 5e). Together, these results confirm that a uniform, ultrathin supramolecular hydrogel coating is successfully formed *in situ via* LLPS, which not only conformally adheres to the complex microtopography of the CFE but also actively promotes early cellular adhesion.

*Enhanced charge-transport characteristics of OP gel*

To probe how the co-assembled architecture influences electrical communication at the coating-electrode interface, we employed light-addressable potentiometric sensor (LAPS) measurements under a 0.15 V bias with 405 nm laser illumination[77]. Spatial photocurrent mapping revealed a well-defined interface between OMP-rich and APA-rich regions. Notably, the interfacial zone where co-assembly occurred exhibited a higher photocurrent response than regions composed of either component alone (Fig. 5c), consistent with a reduction in interfacial impedance. Photocurrent gradient analysis further showed a sharp signal transition across the OMP-APA boundary, indicating pronounced contrast in charge-transport behavior (Fig. 1a, the region highlighted by the dashed box). These results demonstrate that the supramolecular co-assembly not only forms a structurally coherent interface but also creates a pathway for more efficient charge transport compared to the individual components.

*Bulk conductivity and electrically responsive structural ordering of OP gel*

To quantify the macroscopic electrical performance and examine the structural basis of the enhanced charge transport, we measured the bulk conductivity of the OP and APA gels. Prior to characterization, both hydrogels were equilibrated in artificial cerebrospinal fluid (aCSF; conductivity: $1.36 \pm 0.03$ S m$^{-1}$) for 2 hours to achieve a biomimetic ionic balance. The OP gel exhibited a conductivity of $1.72 \pm 0.05$ S m$^{-1}$, exceeding that of the APA gel ($1.46 \pm 0.06$ S m$^{-1}$) (Supplementary Fig. 11). Both values surpass the threshold for conductive biomaterials ($> 0.1$ S m$^{-1}$)[78], confirming the electrically functional nature of the co-assembled hydrogel. Notably, the conductivity of OP gel was higher than that of the aCSF alone, suggesting that the gel scaffold provides a synergistic contribution to the inherent ionic conductivity. We then investigated whether the fibrous network could structurally adapt to electrical cues mimicking neural activity. Electrical stimuli of varying intensities (1-50 mV) and durations (10-600 s) were applied, after which fibre morphology was observed by SEM and fibre orientation was quantified using ImageJ. Increasing stimulation progressively promoted fibre alignment and structural uniformity (Fig. 5f), while unstimulated samples showed a disordered fibre arrangement (Fig. 2e, left). As increased nanofibre alignment within hydrogels is known to enhance charge-transport efficiency by reducing structural disorder[58], these observations indicate that the co-assembled OP gel undergoes an electrically triggered disorder-to-order transition (Fig. 1a, bottom). Collectively, these results demonstrate that OP gel achieves intrinsic conductive-filler-free charge-transport capability through co-assembly and can undergo an electrically triggered disorder-to-order transition, which contributes to enhanced interfacial charge transfer.

**6. *In vivo* tissue integration and anti-inflammatory/scarring performance of OP-CFE**

*Implantation feasibility and preservation of vascular integrity*

To assess whether the OP-gel coating affects the surgical implantation process and acute tissue trauma, OP-CFE and bare CFE were implanted into the mouse M1 motor cortex through a 1-mm cranial burr hole to a depth of 1 mm (Fig. 6a). Cerebral blood flow (CBF) around the implantation site was evaluated 24 h post-surgery using laser speckle contrast imaging (LSCI). The results showed no significant vascular damage or perfusion deficits around either implant type compared with adjacent undisturbed tissue (Fig. 6b). These findings indicate that the microscale electrode design combined with the soft OP-gel interface minimizes acute vascular disruption, confirming the procedural feasibility and tissue-sparing nature of the OF-CFE.

*Early anti-inflammation and anti-scarring effect of OP-CFE in vivo*

To evaluate the ability of the OP-gel coating to modulate the early tissue response, brain tissues surrounding implanted OP-CFE and bare CFE were harvested at 8 hours, 24 hours, and 1 week post-implantation, followed by sectioning and immunofluorescence staining. Confocal laser scanning microscopy (CLSM) revealed distinct differences in local inflammation and glial responses (Fig. 6c-g). At all time points, a significantly higher density of iba-1$^+$ microglia was observed adjacent to bare CFE, indicating pronounced microglial activation and clustering around the uncoated implants while OP-CFE elicited a markedly attenuated microglial response. Furthermore, after 1 week, OP-CFE also provoked a substantially reduced astrocytic reaction compared with bare CFE, as evidenced by GFAP staining. The intensity and spatial spread of GFAP$^+$ signals were significantly lower around OP-CFE, suggesting suppressed astrocytic activation and glial-scar formation. Together, these results demonstrate that the OP-gel coating effectively mitigates acute microglial activation and chronic astrogliosis, thereby favorably modulating the peri-implant inflammatory microenvironment and reducing the risk of immune rejection and glial encapsulation to achieve a seamless neural integration (Fig. 1c, bottom).

**7. Electrochemical and electrophysiological performance evaluation of OP-CFE**

*In vitro electrochemical characterization of OP-CFE*

To assess the influence of OP-gel modification on electrode interfacial properties, we performed electrochemical impedance spectroscopy (EIS) on bare CFE and OP-CFEs with two different coating thicknesses, OP-CFE-4.6 (~4.6 μm) and OP-CFE-3.0 (~3.0 μm) (Supplementary Fig. 12). In the high-to-mid-frequency range, the bare CFE exhibited the largest semicircle diameter, corresponding to a relatively high charge-transfer resistance ($R_{ct}$)[79]. Coating with the OP gel markedly reduced the semicircle diameter for both OP-CFE-4.6 and OP-CFE-3.0, indicating substantially altered interfacial charge-transfer characteristics, which is likely attributable to the intrinsic charge-transport capability of the OP fibres together with the increased electrochemically active surface area[80]. Notably, the extent of impedance reduction was strongly dependent on the thickness of the OP-gel layer. OP-CFE-3.0 showed the smaller semicircle and thus the lower $R_{ct}$ compared with OP-CFE-4.6. This enhanced interfacial performance is likely attributable to an effect of reduced diffusion length at the electrode-electrolyte interface enabled by the ultrathin OP coating. In the low-frequency region, OP-CFE-3.0 displayed a more pronounced inclined tail compared with both CFE and OP-CFE-4.6, suggesting reduced mass-transport impedance, an advantageous feature for neural signal acquisition. In conclusion, these results demonstrate that ultrathin OP-gel coating significantly improves interfacial electrochemical properties, with coating thickness being a critical parameter that governs both charge-transfer and mass-transport behavior at the bioelectronic interface.

*In vivo electrophysiological performance of OP-CFE*

To evaluate the functional recording performance of OP-CFEs in a living system, we implanted both OP-CFEs and bare CFEs in the mouse motor cortex and recorded local field potentials (LFPs) under baseline conditions and during chemically-induced (4-Aminopyridine, 4-AP) seizures[40](experimental design, Fig. 6h). Under baseline conditions, LFPs acquired with OP-CFEs exhibited signal amplitudes across all frequency bands similar to those recorded with bare CFEs (Fig. 6i, left), confirming that the coating does not impair fundamental signal acquisition. During seizure-like episodes, both electrode types reliably captured the expected spectral increases in β- and γ-band power, with no statistically significant differences in frequency-domain metrics between OP-CFEs and bare CFEs (Fig. 6i, right). Collectively, these *in vivo* recordings show that OP-CFE maintain high-fidelity neural signal acquisition across both physiological and high-activity pathological states, fulfilling the core functional requirement for neural interfaces.

**Conclusion**

In this study, we demonstrate the rational design of a supramolecular co-assembly system based on natural OMP and the custom APA to engineer a multifunctional neural interface. This design leverages the intrinsic bioactivity of OMP and the structural programmability of APA to create an OP-gel coating that adaptively responds to the physiological microenvironment. Notably, the OP gel undergoes adaptive fibre architecture and orientation transitions triggered by pH shifts and electrical stimulation, which enhances bioadhesion and charge-transport efficiency *in situ* without extra adhesive moieties and conductive fillers. The *in-situ* LLPS-driven fabrication allows for an ultra-thin (2.8 ± 0.3 μm) conformal coating that maintains high-fidelity signal recording while effectively suppressing neuroinflammation and glial scarring in a mouse cortical model. The main advantages include: (1) molecular and supramolecular design of the system; (2) bioengineering of environment-responsive materials that adapt their hierarchy to physiological cues; (3) facile *in situ* biofabrication *via* LLPS-mediated diffusion-reaction for conformal electrode coating; (4) practical functional validation demonstrating reduced neuroinflammation and stable electrophysiological recording. Collectively, these findings establish a biomimetic framework for constructing seamless bioelectronic interfaces *via* supramolecular co-assembly, offering a versatile platform for advancing chronic neural prosthetics and neuroscience research.

**Methods**

*Synthesis and characterization of PAs*

APA, FITC-APA, AE molecules were provided by GL biochem (China). The sequences and molecular weights of the APA and AE are shown in Figure 2a. PAs were designed as previously reported[81] and synthesized *via* solid-phase chemical synthesis[48].

*Extraction of OMP*

Fresh okra pods were purchased from a local market in Wuhan. After washing and seed removal, the pods were cut into pieces approximately 1 cm² in size. The pieces were mixed with deionized water at a 1:10 (w/v) ratio and stirred continuously at 70 °C for 3 hours. Insoluble impurities were removed by filtration. Subsequently, absolute ethanol was added to the filtrate at a 1:4 (v/v) ratio, followed by thorough stirring and overnight precipitation at 4 °C. The resulting precipitate was collected, redissolved in deionized water, and subjected to a second ethanol precipitation step. The final precipitate was collected, dissolved again, and lyophilized to obtain the crude OMP product[82]. Then crude OMP was decolorized by resin adsorption and thoroughly deproteinized by the Sevage method to obtain the purified OMP product.

*Monosaccharide composition analysis*

5 mg of OMP was hydrolyzed with 2 M Trifluoroacetic acid at 121 °C for 2 h in a sealed tube and dried with nitrogen. Then, methanol was added to wash before being blow-dried, which was repeated 2-3 times. The residue was re-dissolved in deionized water and filtered through 0.22 μm microporous filtering film for measurement. The sample hydrolysates were analyzed by high-performance anion-exchange chromatography (HPAEC) using a Dionex CarboPac PA20 anion-exchange column (150 × 3.0 mm, 10 μm; Thermo Fisher Scientific, US) and a pulsed amperometric detector (PAD). Data were acquired on the Dionex ICS 5000+ system, and processed using Chromeleon 7.2 CDS (Thermo Fisher Scientific, US).

*Molecular weight determination*

The OMP sample was dissolved in 0.1 M $NaNO_3$ aqueous solution containing 0.02% $NaN_3$ at the concentration of 1 mg/mL and filtered through a filter of 0.45 μm pore size. The homogeneity and molecular weight of various fractions were assessed using Size Exclusion Chromatography coupled with Multi-Angle Light Scattering and Refractive Index detection (SEC-MALLS-RI). The weight-average molecular weight (Mw), number-average molecular weight (Mn), and polydispersity index (Mw/Mn) of the different fractions in a 0.1 M $NaNO_3$ aqueous solution containing 0.02% sodium azide ($NaN_3$) were measured utilizing a DAWN HELEOS-II laser photometer (Wyatt Technology Co., US). Data were acquired and processed using ASTRA6.1 (Wyatt Technology).

*Sample preparation of OP gel*

APA was dissolved in deionized water to a final concentration of 1 wt%. OMP was dissolved in normal saline to a final concentration of 0.5 wt%. APA solution was injected into OMP solution. The mixture was then left undisturbed at 4 °C for 1 hour to allow complete gel formation. The resulting OP gel was collected for subsequent use.

*Sample preparation of APA gel*

APA was dissolved in deionized water to a final concentration of 1 wt%. APA solution was injected into normal saline solution. The mixture was then left undisturbed at 4 °C for 1 hour to allow complete gel formation. The resulting APA gel was collected for subsequent use.

*Confocal microscopy*

The interaction and localization of OMP and APA was probed using laser scanning confocal microscopy (Nikon, Japan). OMP (0.5%) was dissolved in a normal saline solution of Thermo Scientific™ Fluoro-Max R25 (diluted 100-fold) and FITC-APA (1 wt%) was dissolved in deionized water. All solutions were incubated for 20 min at room temperature and protected from light. The OP gel was fabricated with 60 μL OMP solution and 20 μL APA solution in PDMS thin film-coated glass slide as previously described. Images were acquired at laser wavelengths of 542 and 488 nm which correspond to the excitation wavelength of Thermo Scientific™ Fluoro-Max R25 and FITC-APA, respectively. Images were further processed using ImageJ.

*Dynamic light scattering (DLS)*

Particle size distributions of OMP and APA were measured using a Nanolink instrument (linkoptik, China). Samples were diluted to a concentration of 0.05 wt% with ultrapure water. Subsequently, the pH of each suspension was adjusted to the desired values using dilute hydrochloric acid or sodium hydroxide solutions prior to measurements.

*Electrophoretic light scattering (ELS)*

Zeta potentials of OMP (0.01 wt%) and PAs (0.01 wt%) solutions were measured at 25 °C using a

NanoLink zeta potential analyzer (linkoptik, China). Samples were prepared in ultrapure water and equilibrated for 30 min at the measurement temperature.

*Scanning electron microscopy (SEM)*

Samples of OMP, OP gel, and OP gel with cells were fixed with 4% paraformaldehyde (PFA, Servicebio, China) for 10 min and dehydrated using an ethanol gradient (50%, 70%, 80%, 90%, 96%, 100%) at room temperature. Then the samples were subjected to critical point drying (HARVENT, China). Dried samples were coated with gold and imaged using GeminiSEM 300 (ZEISS, Germany). The analysis of fibre diameter and orientation was performed using ImageJ.

*Transmission electron microscopy (TEM)*

OMP, APA and OMP-APA co-assembly solutions were deposited onto 200-mesh carbon-coated TEM grids for 5 min, followed by staining with 2 wt% uranyl acetate for 30 s. Grids were rinsed with ultrapure water for 30 s and air-dried at room temperature for 24 h. Bright-field imaging was performed on an HT7800 TEM (HITACHI, Japan) operating at 80 kV.

*Atomic force microscopy (AFM)*

OMP, APA and OMP-APA co-assembly solutions were deposited onto a freshly cleaved mica substrate and allowed to adsorb for 5 minutes. Excess solution was removed by rinsing the mica surface with distilled water and the substrate was dried under a gentle stream of nitrogen gas. Tapping-mode AFM imaging was conducted in air at ambient conditions using a Jupiter AFM (Oxford Instruments, UK).

The OP-gel films were deposited on mica substrates. Force-indentation curves were acquired in liquid environment (different pH conditions) using the force spectroscopy mode on a Jupiter AFM. A colloidal AFM probe was employed to approach and retract from the sample surface. The force curves were subsequently fitted and analyzed using the Johnson-Kendall-Roberts (JKR) contact model within the Asylum Research Software (version 19.10.67).

*Small-angle neutron scattering (SANS)*

The APA was dissolved in $D_2O$ at 1 wt%, and the OMP was dissolved in $D_2O$ at 0.5 wt% (NaCl, 0.9%). OMP (0.5 wt%)-APA (1 wt%) mixed solutions (different pH conditions) were prepared for measurement. Measurements were performed using the SANS instrument (The BL01 at China Spallation Neutron Source, China). Solutions (1 mL) of individual components were loaded into 1-mm-path-length hellma quartz cuvettes, while the mixture was prepared in a demountable 1-mm-path-length cuvette. Cuvettes were mounted in aluminum holders within a sealed, computer-controlled sample chamber maintained at 25 °C. Each measurement took approximately 30 min. All scattering profiles were normalized to sample transmission, background-subtracted using a $D_2O$-filled quartz cell, and detector linearity and efficiency were corrected using the instrument-specific software. The incident neutrons with wavelength of 1.1-9.8 Å were defined by a double-disc bandwidth chopper, which is collimated to the sample by a pair of apertures. The sample-to-detector distance was set to 5 m with an 8 mm sample aperture. The two-dimensional $^3$He tubes array detector allows to cover a wide Q range from 0.004 Å$^{-1}$ to 0.7 Å$^{-1}$. For each sample, including the empty holder and cell, scattering signals were recorded for approximately 60 min. The scattering data were set to absolute unit after normalization, transmission correction and standard sample calibration.

*Rheological test*

Rheological measurements were performed on a Discovery Hybrid Rheometer (TA Instruments, US) using a 20 mm parallel plate. A total of 500 μL of OP gel or APA gel was deposited onto the stage, and the top plate was lowered to a 1000 μm gap. Excess material was removed. The tests included: frequency sweep (0.1-100 rad s$^{-1}$ at 1% strain), time sweep (at a constant shear rate of 1 s$^{-1}$), strain sweep (0.1-100% strain at 1 rad s$^{-1}$), shear rate sweep (0.01-100 s$^{-1}$), and shear recovery (a minute cycles each at high (500%), low (1%) strain at 1 rad s$^{-1}$).

*Cell adhesion test*

Firstly, the OP-gel films were deposited on the newly cleaved mica substrate, and the pH value of the thin films was altered by adding dilute hydrochloric acid (pH 4) and sodium hydroxide (pH 8) onto the thin films for 2 minutes. DiI-fluorescently labelled BV-2 cells were plated in 12-well plates at a concentration of $0.5 \times 10^6$ cells per well and incubated with the films and bare mica substrates for 6 h in cell media at 37℃ with 5% $CO_2$. Post-incubated films and bare mica substrates were washed thoroughly with warm PBS buffer to remove cells that had not strongly adhered to the surface, and then fluorescence images were captured on laser scanning confocal microscopy. At least four representative images were taken for each sample, and adhered cells were counted by ImageJ.

*Cell viability and cytotoxicity testing*

HUVECs were obtained from Procell (China) and cultured in RPMI medium (Gibco, US) supplemented with 10% FBS (Gibco, US) and 1% PS (Gibco, US). BV-2 was obtained from Shanghai Institute of Biochemistry and Cell Biology (China) and cultured in DMEM High-Glucose medium (Gibco, US) supplemented with 10% FBS (Gibco, US) and 1% PS (Gibco, US). HUVECs or BV-2 (10,000 cells per well) were seeded in the well of a 24-well plate, while the OP gel or APA gel were placed in the Transwell insert (8.0 μm pore size, Corning, US) and co-cultured for 7 days. Each gel had a fixed volume of 50 μL. Cytotoxicity was evaluated using a Cell Counting Kit 8 (Abcam, UK) according to the assay instructions. Plates were subsequently incubated for 2 h at 37 °C and absorbance was measured at 450 nm using a microplate reader (BioTek, US). Both cells (5,000 cells per well) were seeded on the surface of OP gel and APA gel (50 μL per well in the 96-well plate) to facilitate 3D growth for 7 days. Cell viability was assessed using Live/Dead Cell Double Staining Kit (Sigma, US) on days 1, 3, and 7. A mixture of propidium iodide (PI) and Calcein AM in PBS was applied for 30 min at 37 °C. Fluorescence images were captured on laser scanning confocal microscopy.

*Phalloidin staining*

Phalloidin staining was performed at 1, 3, and 7 days post-cell seeding. At each time point, OP gels with cells were washed three times with PBS and fixed with 4% PFA for 30 min at room temperature, followed by permeabilization with 0.5% Triton X-100 for 20 minutes. After thorough washing with PBS, samples were incubated with Alexa Fluor 594-phalloidin (Thermo Fisher Scientific, US) diluted in PBS for 60 minutes, protected from light. Cell nuclei were counterstained with DAPI for 10 minutes. Samples were washed with PBS and imaged using a confocal microscope.

*Establishment of the BV-2 neuroinflammation model*

A standardized *in vitro* neuroinflammation model was established using the BV-2 cells stimulated with lipopolysaccharide (LPS). BV-2 cells were seeded into 24-well plates at a density of $20 \times 10^4$ cells per well. After allowing sufficient time for cell adhesion, the medium was replaced with serum-free medium for 12 h. Subsequently, the serum-free medium was replaced with fresh serum-free medium containing 1 μg/mL LPS (Sigma, US). The cells were stimulated with LPS for 24 hours under standard culture conditions (37°C, 5% $CO_2$).

*Detection of extracellular NO release and ROS production*

Concurrently with LPS stimulation, a Transwell insert pre-loaded with OP gel was placed into the well containing BV-2 cells as the experimental group. Control groups included untreated cells and cells treated with LPS alone. After 24 hours, the culture supernatants were collected. The concentration of nitric oxide (NO) in the supernatant was quantified using a Nitric Oxide Assay Kit (Nitrate reductase method, Nanjing Jiancheng Bioengineering Institute, China) according to the manufacturer's instructions. Absorbance was measured at 550 nm using a microplate reader. Cells were first stained with Hoechst 33342 (Solabrio, China) for nuclei labeling. Subsequently, intracellular ROS levels were detected using the fluorescent probe DCFH-DA from a ROS Assay Kit (Solabrio, China). And a positive control was established by treating untreated cells with Rosup for 20 minutes following DCFH-DA incubation. Fluorescence imaging was performed using a laser scanning confocal microscope. The mean fluorescence intensity of DCF (oxidized product) was quantified using ImageJ.

*Conductivity test*

Cylindrical samples of OP gel and APA gel, each with a cross-sectional diameter of 10 mm and a thickness of 3 mm, were fabricated in a 48-well plate. To mimic the ionic equilibrium of the physiological environment, both types of hydrogel discs were immersed in artificial cerebrospinal fluid (aCSF, Sigma, US) for 2 hours. Following immersion, excess surface aCSF was carefully removed by blotting with filter paper. The electrical resistivity of the hydrated hydrogel samples was then measured using a four-point probe resistivity meter (Jingge, China). Measurements were taken at multiple distinct points on each sample to ensure reproducibility. In parallel, the conductivity of the aCSF solution was measured using a calibrated conductivity meter (OHAUS, US).

*Light-addressable potentiometric sensor (LAPS) test*

The LAPS system[77] employed a 405 nm laser diode (maximum power: 10 mW; modulation frequency: 1 kHz) as the light source. The beam was focused onto the sensor surface using a 10× objective lens. The $Si-SiO_2-Si_3N_4$ wafer served as the sensor chip, which was integrated into a custom fluidic measurement cell mounted on a motorized XYZ translation stage for precise spatial positioning. A standard three-electrode configuration was used, comprising a platinum counter electrode and an Ag/AgCl reference electrode. Photocurrent signals were acquired and processed using a lock-in amplifier at room temperature. The functional surface of sensor chip was initially covered with 1.0 wt% APA solution. Subsequently, the 0.5 wt% OMP solution was injected into the APA solution and allowed to incubate for 10 minutes to facilitate interfacial co-assembly. Finally, a ten-fold diluted PBS buffer was introduced to completely fill the measurement cell and establish a stable electrochemical environment. Then the photocurrent was monitored in real time.

*Sample preparation of OP-CFE*

Precursor solutions were prepared using sterile normal saline for OMP (0.5 wt%) and sterile deionized water for APA (1 wt%). Carbon fibre electrodes (CFEs), featuring a tip diameter of 7 μm, were sterilized overnight using a UV-ozone cleaner prior to coating. The OP-CFE functionalization was performed via a layer-by-layer dip-coating technique at ambient temperature. Briefly, the CFE tips were alternately immersed in the APA solution for 5 min and the OMP solution for 5 min. This deposition cycle was repeated to tune the coating thickness; unless otherwise specified, 5 cycles were employed to achieve a uniform thickness of 3.0 ± 0.1 μm. The morphology and precise thickness of the resulting OP-gel layer were characterized using SEM and quantified using ImageJ. To ensure stability and sterility, all OP-CFEs were freshly prepared before each experiment and stored with the tips submerged in sterile normal saline until use.

*Electrochemical impedance spectroscopy (EIS) test*

Electrochemical impedance spectra were acquired using a CHI 660E electrochemical workstation (Shanghai Chenhua Instrument, China) with a conventional three-electrode configuration. An Ag/AgCl electrode served as the reference electrode. Measurements were performed in a potassium ferricyanide solution (0.1 M KCl + 5mM $K_3[Fe(CN)_6]/K_4[Fe(CN)_6]$). The impedance spectra of bare CFE and OP-CFE with two distinct coating thicknesses (~4.6 μm and ~3.0 μm) were recorded across a frequency range of 100 kHz to 0.1 Hz.

*Animal experiments (mouse subcutaneous implantation model)*

Animal procedures were approved by the Institutional Animal Care and Use Committee of Huazhong University of Science and Technology ([2025] IACUC Number:5022). BALB/c mice were anesthetized *via* intraperitoneal injection of tribromoethanol (250 mg/kg, Nanjing Aibei Biological Technology Co., Ltd., China). OP gel was subcutaneously grafted over the dorsum of the mice. The control group consisted of APA gel. Each gel had a fixed volume of 0.25 mL. Prior to grafting, the dorsal skin of the mice was sterilized. OP gel and APA gel were injected separately into the respective subcutaneous cavities created above the muscle layer on bilateral dorsal sides. After surgery, the mice were placed in an incubator until full recovery from anesthesia. The implantation site and major organs were harvested at 2 weeks post-grafting (n = 3 per group) following euthanasia.

*Immunohistology staining*

Subcutaneous grafts were fixed with 4% PFA and processed for paraffin embedding. Sections were deparaffinized, rehydrated, and blocked with 3% BSA for 30 min at room temperature. Primary antibody (ab5694, Abcam, UK) against α-SMA (for capillaries) was applied overnight at 4 °C. After washing, secondary fluorophore-conjugated antibodies (Service, China) were added for 50 min at room temperature. DAPI (Beyotime, China) was used for nuclear staining. Samples were visualized using a confocal microscope. All image analysis was performed using ImageJ.

*Histological staining*

Major organs of mice were harvested and fixed in 4% PFA overnight at 4 °C. The tissue specimen was sectioned in the cross-sectional or longitudinal direction and stained with haematoxylin and eosin (Servicebio, China). Samples were visualized using an optical microscope (Nikon, Japan).

*Animal experiments (mouse cortical implantation model)*

Mice were deeply anesthetized *via* tribromoethanol and the head was fixed in a stereotaxic frame, and the scalp was shaved and sterilized. A midline incision was made to expose the skull. Bilateral craniotomies (1 mm in diameter) were created above the primary motor cortex (M1) using a precision dental drill. The underlying dura mater was carefully punctured and reflected. A bare CFE was implanted into the left cortical hemisphere to a depth of 1.0 mm from the cortical surface, while an OP-CFE was implanted to the same depth in the right hemisphere. Electrodes were secured to the skull using dental acrylic cement, and the scalp incision was sutured. At 8 h, 24 h, and 1 week post-implantation, mice (n = 3 per group per time point) were deeply anesthetized and the electrodes were carefully extracted. Animals were then transcardially perfused with ice-cold PBS followed by 4% PFA. The brain was dissected, post-fixed in 4% PFA at 4 °C overnight, and cryoprotected sequentially in 20% and 30% sucrose solution for histopathological analysis.

*Laser speckle contrast imaging (LSCI)*

24 h after the initial electrode implantation surgery, mice were anesthetized *via* tribromoethanol. Following the careful surgical re-exposure of the skull, the dental acrylic cap and the implanted electrodes were meticulously removed. The cranial surface was then kept moist with sterile physiological saline (0.9% NaCl). The animal was positioned under the LSCI camera (RWD, China), and the region of interest (ROI) was defined to encompass the original cranial window over the primary motor cortex (M1), specifically targeting the site of prior electrode implantation. After confirming stable physiological parameters, a baseline recording of cerebral blood flow was acquired. The cerebral blood flow (CBF) maps were performed using the system's proprietary software.

*Immunohistology staining and quantitative analysis*

The brain tissues were sectioned coronally (20 μm thickness) perpendicular to the electrode track using a freezing microtome (ThermoFisher Scientific, US). Sections were immunofluorescently labeled with primary antibodies against iba-1 (Abcam, UK) for microglia and GFAP (Oasis, China) for astrocytes. Nuclei were counterstained with DAPI. Slides were mounted with ProLong Gold and imaged on a laser scanning confocal microscope. Fluorescence intensity was quantified as a function of distance from the implantation center using ImageJ. The center of the electrode track was defined as $x = 0$ μm. A concentric analysis was performed by dividing the adjacent region (up to 160 μm from the center) into circular contours of 20 μm increments, and the average intensity within each contour was calculated. For non-implanted control tissue, the center of the image field was set as $x = 0$ μm.

*Animal experiments (intracranial local field potential recording)*

Mice (n = 6) were anesthetized *via* tribromoethanol and craniotomies (~1 mm diameter) were drilled above the left M1 cortex (three holes spaced 1 mm apart) and bilaterally over the cerebellum to expose the dura. Two stainless-steel screws were inserted into the cerebellar burr holes to serve as reference and ground electrodes. Both OP-CFE and bare CFE were implanted into adjacent sites within the left M1 cortex. All electrodes and wires were secured with dental acrylic. Following surgery, the electrodes were connected to a pre-amplifier for local field potential (LFP) acquisition. 4-AP (15 mM, 500 nL, Sigma, US) was microinjected through the third cranial hole, and LFPs were recorded again 15 min later. Raw signals were band-pass filtered and analyzed in the frequency domain.

*Statistical analysis*

Statistical analysis was performed using unpaired t test, one-way or two-way ANOVA using GraphPad Software (Prism v.10). P values were used to determine significant differences (ns at $P > 0.05$, *$P < 0.05$, **$P < 0.01$, ***$P < 0.001$, and ****$P < 0.0001$).

**Supplemental Information Description**

All data are available in the main text or the supplementary materials.

**Acknowledgements**

This work was supported by the National Natural Science Foundation of China (NSFC) for the Excellent Young Scientists Fund (Overseas, 0214530013), the Young Scientists Fund of the NSFC (82302837), the China Aerospace Science and Technology Corporation (0231530004), the Strategic Partnership Research Funding (HUST-Queen Mary University of London, No. 2022-HUST-QMUL-SPRF-07), the Natural Science Foundation of Hubei Province, China (Grant No. 2024AFB676). We thank the Medical Subcentre of Huazhong University of Science and Technology (HUST) Analytical & Testing Centre. This project was approved by the China Spallation Neutron Source (CSNS) under the grant number P0123122900034 and we thank the staff members of the Small Angle Neutron Scattering (https://cstr.cn/31113.02.CSNS.SANS) at the CSNS (https://cstr.cn/31113.02.CSNS) for providing technical support and assistance in data collection. Parts of figures were drawn by Figdraw.

**Author contributions**

Y.W., D.Z., J.W., T.L. and Y.G. conceived the project. T.L. and Y.G. carried out the experiments. Y.W., D.Z. and J.W. supervised the study. S.S. performed biological characterization. Q.Y. performed fabrication of carbon fibre electrodes and light-addressable potentiometric sensor experiments. Z.Y. performed the OP gel co-assembling characterization. Z.H. performed animal experiments. D.Y. conducted the SANS and analysed the data. Y.K. and H.Y. assisted with SANS.

**Competing interests**

The authors declare no competing interests.

**Correspondence and requests for materials** should be addressed to Yuanhao Wu, Dewen Zhang and Jiecong Wang.

**References**

1. Frank, J.A., Antonini, M.J. & Anikeeva, P. Next-generation interfaces for studying neural function. *Nat Biotechnol* **37**, 1013–1023 (2019).
2. Lebedev, M.A. & Nicolelis, M.A. Brain-Machine Interfaces: From Basic Science to Neuroprostheses and Neurorehabilitation. *Physiol Rev* **97**, 767–837 (2017).
3. Cagnan, H., Denison, T., McIntyre, C. & Brown, P. Emerging technologies for improved deep brain stimulation. *Nat Biotechnol* **37**, 1024–1033 (2019).
4. Santhanam, G., Ryu, S.I., Yu, B.M., Afshar, A. & Shenoy, K.V. A high-performance brain-computer interface. *Nature* **442**, 195–198 (2006).

5. Yang, X. et al. Kirigami electronics for long-term electrophysiological recording of human neural organoids and assembloids. *Nat Biotechnol* **42**, 1836–1843 (2024).
6. Sui, Y. et al. Deep brain-machine interfaces: sensing and modulating the human deep brain. *Natl Sci Rev* **9**, nwac212 (2022).
7. Apollo, N.V. et al. Gels, jets, mosquitoes, and magnets: a review of implantation strategies for soft neural probes. *J Neural Eng* **17**, 041002 (2020).
8. Zhang, E.N. et al. Mechanically matched silicone brain implants reduce brain foreign body response. *Advanced Materials Technologies* **6**, 2000909 (2021).
9. Karumbaiah, L. et al. The upregulation of specific interleukin (IL) receptor antagonists and paradoxical enhancement of neuronal apoptosis due to electrode induced strain and brain micromotion. *Biomaterials* **33**, 5983–5996 (2012).
10. Boufidis, D., Garg, R., Angelopoulos, E., Cullen, D.K. & Vitale, F. Bio-inspired electronics: Soft, biohybrid, and "living" neural interfaces. *Nat Commun* **16**, 1861 (2025).
11. Chen, D. et al. An Ultra-Flexible Neural Electrode with Bioelectromechanical Compatibility and Brain Micromotion Detection. *Adv Healthc Mater*, e03101 (2025).
12. Salatino, J.W., Ludwig, K.A., Kozai, T.D.Y. & Purcell, E.K. Glial responses to implanted electrodes in the brain. *Nat Biomed Eng* **1**, 862–877 (2017).
13. Trotier, A. et al. Micromotion Derived Fluid Shear Stress Mediates Peri-Electrode Gliosis through Mechanosensitive Ion Channels. *Adv Sci (Weinh)* **10**, e2301352 (2023).
14. Polikov, V.S., Block, M.L., Fellous, J.M., Hong, J.S. & Reichert, W.M. In vitro model of glial scarring around neuroelectrodes chronically implanted in the CNS. *Biomaterials* **27**, 5368–5376 (2006).
15. Wang, L. et al. Bioaugmented design and functional evaluation of low damage implantable array electrodes. *Bioact Mater* **47**, 18–31 (2025).
16. Shen, K., Chen, O., Edmunds, J.L., Piech, D.K. & Maharbiz, M.M. Translational opportunities and challenges of invasive electrodes for neural interfaces. *Nat Biomed Eng* **7**, 424–442 (2023).
17. Tian, G. et al. Electrostatic Interaction-Based High Tissue Adhesive, Stretchable Microelectrode Arrays for the Electrophysiological Interface. *ACS Appl Mater Interfaces* **14**, 4852–4861 (2022).
18. Lao, J. et al. Intrinsically Adhesive and Conductive Hydrogel Bridging the Bioelectronic-Tissue Interface for Biopotentials Recording. *ACS Nano* **19**, 7755–7766 (2025).
19. Perez-Chirinos, L. et al. Tuning the Dimensionality of Protein-Peptide Coassemblies to Build 2D Conductive Nanomaterials. *ACS Nano* **19**, 16500–16516 (2025).
20. Wang, L. et al. Tough and Functional Hydrogel Coating by Electrostatic Spraying. *Small*, e2408780 (2024).
21. Liang, Q. et al. Electron Conductive and Transparent Hydrogels for Recording Brain Neural Signals and Neuromodulation. *Adv Mater* **35**, e2211159 (2023).
22. Zhang, J. et al. Engineering Electrodes with Robust Conducting Hydrogel Coating for Neural Recording and Modulation. *Adv Mater* **35**, e2209324 (2023).
23. Khan, W.U., Shen, Z., Mugo, S.M., Wang, H. & Zhang, Q. Implantable hydrogels as pioneering materials for next-generation brain-computer interfaces. *Chem Soc Rev* **54**, 2832–2880 (2025).
24. Xue, X.Y. et al. Conductive Hydrogel-Based Neural Interfaces: From Fabrication Methods, Properties, to Applications. *Small Structures*, 2400696 (2025).

25. Zhang, P. et al. Conducting Hydrogel-Based Neural Biointerfacing Technologies. *Advanced Functional Materials*, 2422869 (2025).
26. Lv, S. et al. Long-term stability strategies of deep brain flexible neural interface. *NPJ Flexible Electronics* **9**, 40 (2025).
27. Chu, T. et al. Highly Conductive, Adhesive and Biocompatible Hydrogel for Closed-Loop Neuromodulation in Nerve Regeneration. *ACS Nano* **19**, 18729–18746 (2025).
28. Guo, W.-Y. & Ma, M.-G. Conductive nanocomposite hydrogels for flexible wearable sensors. *Journal of Materials Chemistry A* **12**, 9371–9399 (2024).
29. Chen, J., Liu, F., Abdiryim, T. & Liu, X. An overview of conductive composite hydrogels for flexible electronic devices. *Advanced Composites and Hybrid Materials* **7**, 35 (2024).
30. Zhou, T. et al. 3D printable high-performance conducting polymer hydrogel for all-hydrogel bioelectronic interfaces. *Nat Mater* **22**, 895–902 (2023).
31. Vashist, A. et al. Advances in Carbon Nanotubes-Hydrogel Hybrids in Nanomedicine for Therapeutics. *Adv Healthc Mater* **7**, e1701213 (2018).
32. Kougkolos, G., Golzio, M., Laudebat, L., Valdez-Nava, Z. & Flahaut, E. Hydrogels with electrically conductive nanomaterials for biomedical applications. *J Mater Chem B* **11**, 2036–2062 (2023).
33. Zhang, Y., Gong, M. & Wan, P. MXene hydrogel for wearable electronics. *Matter* **4**, 2655–2658 (2021).
34. Won, D. et al. Digital selective transformation and patterning of highly conductive hydrogel bioelectronics by laser-induced phase separation. *Sci Adv* **8**, eabo3209 (2022).
35. Wang, J. et al. Ultra-High Electrical Conductivity in Filler-Free Polymeric Hydrogels Toward Thermoelectrics and Electromagnetic Interference Shielding. *Adv Mater* **34**, e2109904 (2022).
36. Capito, R.M., Azevedo, H.S., Velichko, Y.S., Mata, A. & Stupp, S.I. Self-assembly of large and small molecules into hierarchically ordered sacs and membranes. *Science* **319**, 1812–1816 (2008).
37. Wu, Y. et al. Disordered protein-graphene oxide co-assembly and supramolecular biofabrication of functional fluidic devices. *Nat Commun* **11**, 1182 (2020).
38. Wang, H., Mills, J., Sun, B. & Cui, H. Therapeutic Supramolecular Polymers: Designs and Applications. *Prog Polym Sci* **148** (2024).
39. Finkelstein-Zuta, G. et al. A self-healing multispectral transparent adhesive peptide glass. *Nature* **630**, 368–374 (2024).
40. Nam, J. et al. Supramolecular Peptide Hydrogel-Based Soft Neural Interface Augments Brain Signals through a Three-Dimensional Electrical Network. *ACS Nano* **14**, 664–675 (2020).
41. Jain, D. et al. Low-Molecular-Weight Hydrogels as New Supramolecular Materials for Bioelectrochemical Interfaces. *ACS Appl Mater Interfaces* **9**, 1093–1098 (2017).
42. Xu, H. et al. An investigation of the conductivity of peptide nanotube networks prepared by enzyme-triggered self-assembly. *Nanoscale* **2**, 960–966 (2010).
43. Tovar, J.D., Rabatic, B.M. & Stupp, S.I. Conducting polymers confined within bioactive peptide amphiphile nanostructures. *Small* **3**, 2024–2028 (2007).
44. Arnold, M.S., Guler, M.O., Hersam, M.C. & Stupp, S.I. Encapsulation of carbon nanotubes by self-assembling peptide amphiphiles. *Langmuir* **21**, 4705–4709 (2005).
45. Okesola, B.O. et al. Covalent co-assembly between resilin-like polypeptide and peptide amphiphile into hydrogels with controlled nanostructure and improved mechanical properties. *Biomater Sci* **8**, 846–857 (2020).

46. Wu, Y. et al. Co-assembling living material as an in vitro lung epithelial infection model. *Matter* **7**, 216–236 (2024).

47. Wu, Y. et al. Disinfector-assisted low temperature reduced graphene oxide-protein surgical dressing for the postoperative photothermal treatment of melanoma. *Advanced Functional Materials* **32**, 2205802 (2022).

48. Su, S. et al. De novo design of alpha-helical peptide amphiphiles repairing fragmented collagen type I via supramolecular co-assembly. *arXiv preprint arXiv:2507.14577* (2025).

49. Agregán, R. et al. Biological activity and development of functional foods fortified with okra (Abelmoschus esculentus). *Crit Rev Food Sci Nutr* **63**, 6018–6033 (2023).

50. Zhu, X.M., Xu, R., Wang, H., Chen, J.Y. & Tu, Z.C. Structural Properties, Bioactivities, and Applications of Polysaccharides from Okra [Abelmoschus esculentus (L.) Moench]: A Review. *J Agric Food Chem* (2020).

51. Xu, Y., Cao, H. & He, J. Research advances in okra polysaccharides: Green extraction technology, structural features, bioactivity, processing properties and application in foods. *Food Res Int* **202**, 115686 (2025).

52. Liu, Y., Ye, Y., Hu, X. & Wang, J. Structural characterization and anti-inflammatory activity of a polysaccharide from the lignified okra. *Carbohydr Polym* **265**, 118081 (2021).

53. Wang, C., Yu, Y.B., Chen, T.T., Wang, Z.W. & Yan, J.K. Innovative preparation, physicochemical characteristics and functional properties of bioactive polysaccharides from fresh okra (Abelmoschus esculentus (L.) Moench). *Food Chem* **320**, 126647 (2020).

54. Kaur, G., Singh, D. & Brar, V. Bioadhesive okra polymer based buccal patches as platform for controlled drug delivery. *Int J Biol Macromol* **70**, 408–419 (2014).

55. Nie, X.R. et al. Structural characteristics, rheological properties, and biological activities of polysaccharides from different cultivars of okra (Abelmoschus esculentus) collected in China. *Int J Biol Macromol* **139**, 459–467 (2019).

56. Gorges, H., Kovalev, A. & Gorb, S.N. Structure, mechanical and adhesive properties of the cellulosic mucilage in Ocimum basilicum seeds. *Acta Biomater* **184**, 286–295 (2024).

57. Zhao, Z., Dong, Z., Chen, C., Peng, J. & Ma, P. Synergistically improving interface behavior by designing physical twisting structure and "rigid-flexible" interface layer on ultra-high molecular weight polyethylene (UHMWPE) fiber surface. *Thin-Walled Structures* **199**, 111805 (2024).

58. Chanda, A., Sinha, S.K. & Datla, N.V. Electrical conductivity of random and aligned nanocomposites: Theoretical models and experimental validation. *Composites Part A: Applied Science and Manufacturing* **149**, 106543 (2021).

59. Won, C. et al. Emerging fiber-based neural interfaces with conductive composites. *Mater Horiz* **12**, 4545–4572 (2025).

60. Chen, Y. et al. Helical peptide structure improves conductivity and stability of solid electrolytes. *Nat Mater* **23**, 1539–1546 (2024).

61. Ing, N.L., Spencer, R.K., Luong, S.H., Nguyen, H.D. & Hochbaum, A.I. Electronic Conductivity in Biomimetic α-Helical Peptide Nanofibers and Gels. *ACS Nano* **12**, 2652–2661 (2018).

62. Grosvirt-Dramen, A. et al. Hierarchical Assembly of Conductive Fibers from Coiled-Coil Peptide Building Blocks. *ACS Nano* **19**, 10162–10172 (2025).

63. Zhang, Z. et al. Supramolecular Structure Enabled Photo-Responsive Charge Transport in Porphyrin-Based Junctions. *Angew Chem Int Ed Engl* **64**, e202508443 (2025).

64. Kelly, P.V., Gardner, D.J. & Gramlich, W.M. Optimizing lignocellulosic nanofibril dimensions and morphology by mechanical refining for enhanced adhesion. *Carbohydr Polym* **273**, 118566 (2021).

65. Fu, H. et al. Supramolecular polymers form tactoids through liquid-liquid phase separation. *Nature* **626**, 1011–1018 (2024).

66. Inostroza-Brito, K.E. et al. Co-assembly, spatiotemporal control and morphogenesis of a hybrid protein–peptide system. *Nature Chemistry* **7**, 897–904 (2015).

67. Zheng, R. et al. Assembly of short amphiphilic peptoids into nanohelices with controllable supramolecular chirality. *Nat Commun* **15**, 3264 (2024).

68. Hu, X. et al. Neutron reflection and scattering in characterising peptide assemblies. *Adv Colloid Interface Sci* **322**, 103033 (2023).

69. Lu, Y.B. et al. Viscoelastic properties of individual glial cells and neurons in the CNS. *Proc Natl Acad Sci U S A* **103**, 17759–17764 (2006).

70. Denk, J. et al. Synergistic enhancement of thermomechanical properties and oxidation resistance in aligned Co-continuous carbon-ceramic hybrid fibers. *Mater Horiz* **11**, 5777–5785 (2024).

71. Xu, J., Zhu, X., Zhao, J., Ling, G. & Zhang, P. Biomedical applications of supramolecular hydrogels with enhanced mechanical properties. *Adv Colloid Interface Sci* **321**, 103000 (2023).

72. Chesler, M. Regulation and modulation of pH in the brain. *Physiological reviews* **83**, 1183–1221 (2003).

73. Zheng, H., Zhang, Z., Cai, S., An, Z. & Huang, W. Enhancing Purely Organic Room Temperature Phosphorescence via Supramolecular Self-Assembly. *Adv Mater* **36**, e2311922 (2024).

74. Lam, C.D. & Park, S. Nanomechanical characterization of soft nanomaterial using atomic force microscopy. *Mater Today Bio* **31**, 101506 (2025).

75. Li, F. et al. Low-intensity pulsed ultrasound stimulation (LIPUS) modulates microglial activation following intracortical microelectrode implantation. *Nat Commun* **15**, 5512 (2024).

76. Villa, J., Cury, J., Kessler, L., Tan, X. & Richter, C.P. Enhancing biocompatibility of the brain-machine interface: A review. *Bioact Mater* **42**, 531–549 (2024).

77. Chen, F. et al. Visualization of electrochemical reactions on microelectrodes using light-addressable potentiometric sensor imaging. *Anal Chim Acta* **1224**, 340237 (2022).

78. Wang, S. et al. 3D culture of neural stem cells within conductive PEDOT layer-assembled chitosan/gelatin scaffolds for neural tissue engineering. *Mater Sci Eng C Mater Biol Appl* **93**, 890–901 (2018).

79. Hallaj, R., Ghafary, Z., Kamal Mohammed, O. & Shakeri, R. Induced ultrasensitive electrochemical biosensor for target MDA-MB-231 cell cytoplasmic protein detection based on RNA-cleavage DNAzyme catalytic reaction. *Biosens Bioelectron* **227**, 115168 (2023).

80. Wei, M. et al. How to Choose a Proper Theoretical Analysis Model Based on Cell Adhesion and Nonadhesion Impedance Measurement. *ACS Sens* **6**, 673–687 (2021).

81. Mata, A., Palmer, L., Tejeda-Montes, E. & Stupp, S.I. in Nanotechnology in Regenerative Medicine: Methods and Protocols 39–49 (Springer, 2011).

82. Liu, J. et al. Structure characterisation of polysaccharides in vegetable "okra" and evaluation of hypoglycemic activity. *Food Chemistry* **242**, 211–216 (2018).

we

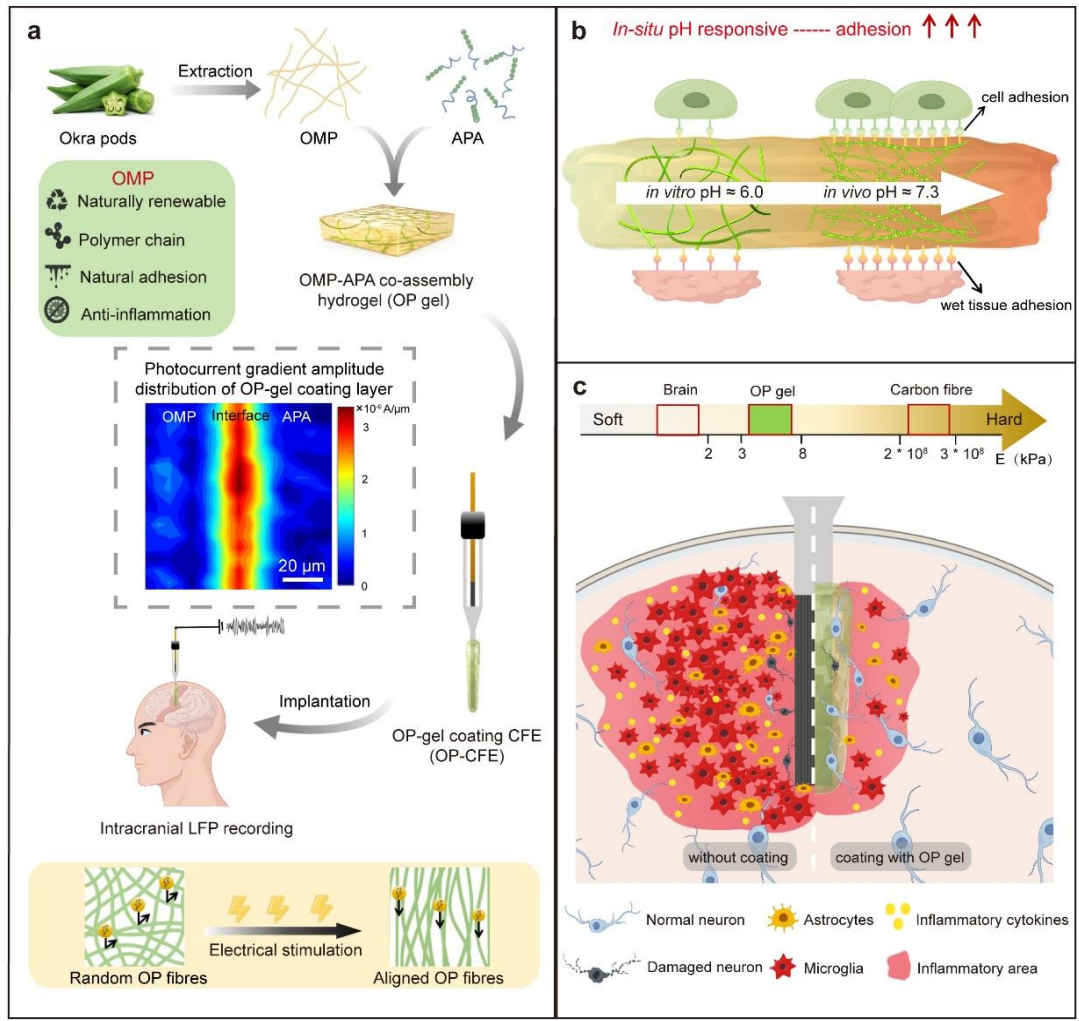

**Fig. 1 | Rationale for co-assembling system and its neural electrode coating applications. a**. The illustration of okra mucilage polysaccharide (OMP) and α-helical peptide amphiphiles (APA), and flowchart of the steps used to prepare the OMP-APA co-assembly hydrogel (OP gel) as coating layer onto the carbon fibre electrodes (CFE) and the application of OP-gel-coated electrodes (OP-CFE). Photocurrent gradient analysis showed a sharp signal transition across the OMP-APA co-assembling interface, indicating pronounced contrast in charge-transport behavior (the region highlighted by the dashed box). Illustration of electrically enhanced fibre alignment and charge transport in OP gel (bottom). **b**. Illustration of the structural adaptation of OP-gel fibres to the *in vivo* pH environment for enhanced bioadhesion. **c**. Comparative illustration of the mechanical modulus among OP gel, carbon fibre and brain tissue, highlighting the superior compatibility of OP gel with neural tissue (top). Comparative illustration of the cellular and microenvironmental responses following the implantation of bare CFE versus OP-CFE, demonstrating enhanced biointegration at the electrode-neural interface mediated by the OP-gel coating (bottom).

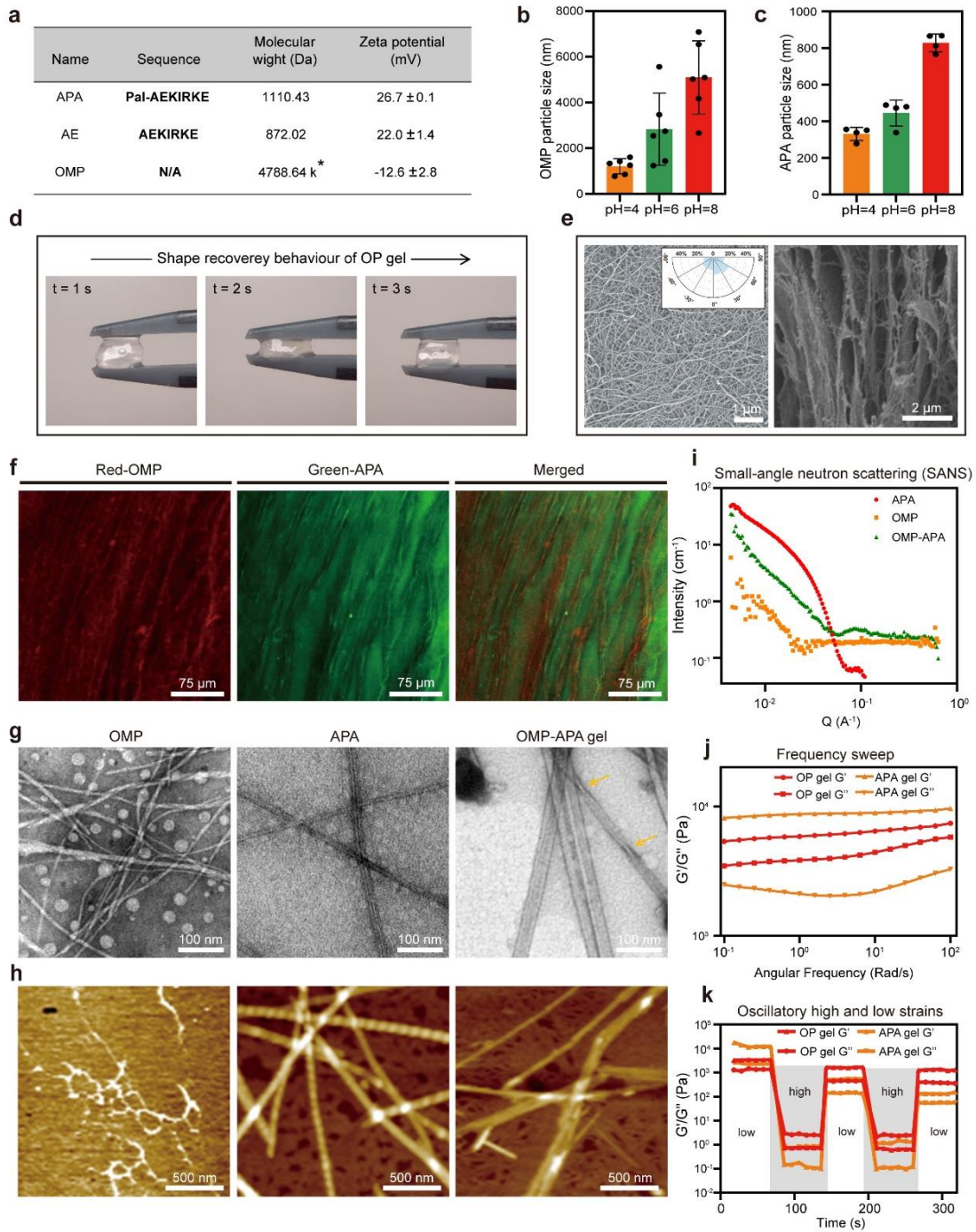

**Fig. 2 | Characterization of the OMP-APA co-assembly system. a**. Table summarizing the key information of the APA and AE molecules and OMP used in the study (*peak molecular weight). **b & c**, DLS analysis revealing both OMP and APA undergo systematic, pH-dependent self-assembly behaviours. **d**, Recovery of OP gel after mechanical pinching. **e**, SEM images of the surface and the cross-sectional views of the OP gel. **f**, Confocal microscopy (green: APA, red: OMP) corroborating the interaction between the OMP and APA (the strongly overlapping signals of the two components parallel to the interface). **g & h**, TEM and AFM images of the system. **i**, Small-angle neutron scattering (SANS) patterns of 151-Å radius OMP (yellow square) and 59-Å radius APA (red circle) cylindrical nanofibres, and a resulting uniform cylindrical nanofibre of OMP-APA co-assembly with

38-Å radius (green triangle). **j**, Shear sweeps from 0.01 to 100 s$^{-1}$ of OP gel and APA gel. **k**, Oscillatory high and low strains (white regions represent 1% strain and grey regions represent 500% strain) of OP gel and APA gel.

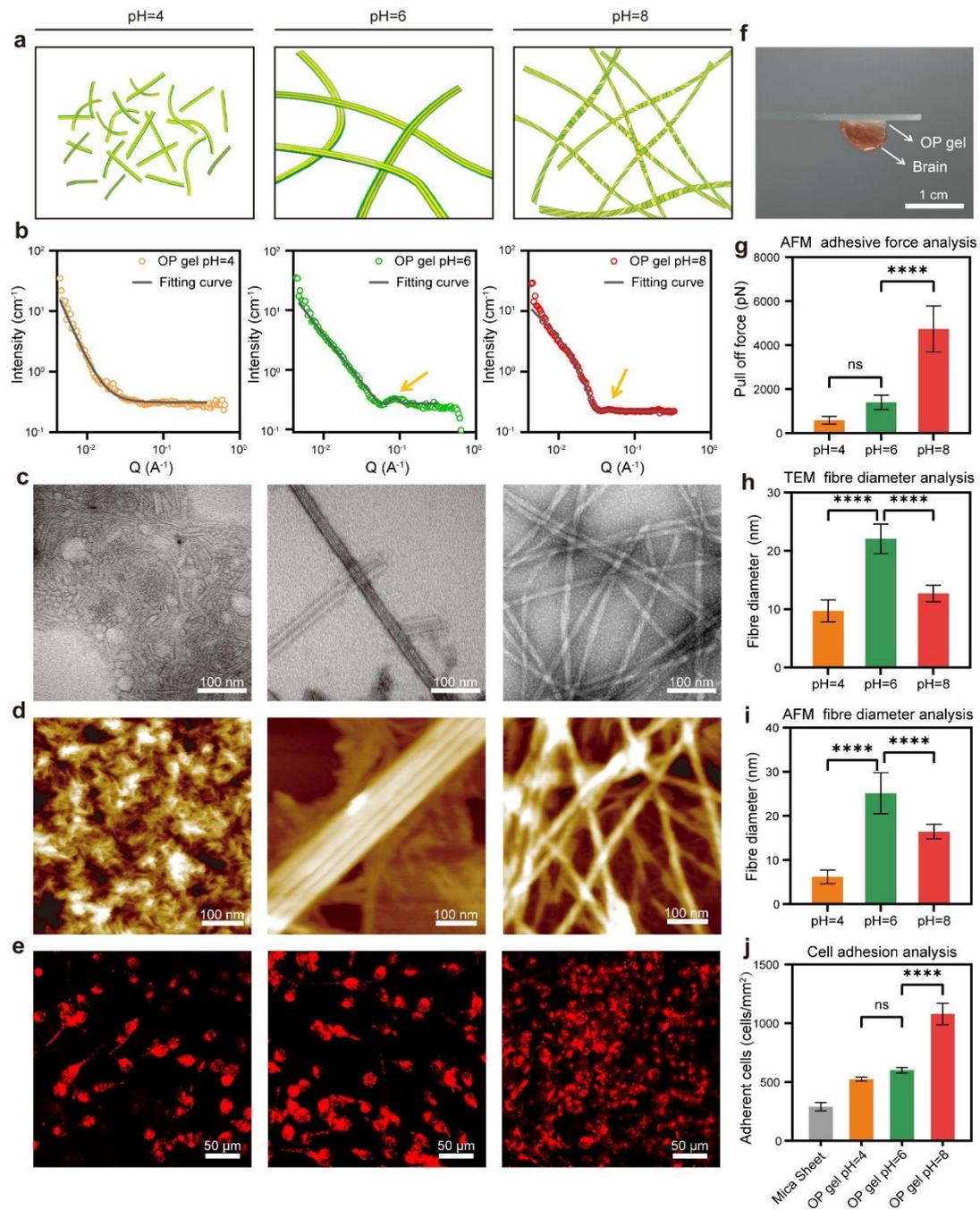

**Fig. 3 | *In situ* pH responsive adaption of the OMP-APA co-assembly system. a**. Illustration of the OMP-APA co-assembly fibre structures at pH 4, 6, and 8. **b**. SANS patterns demonstrating the OP gel undergoes pronounced, pH-driven reorganization of its co-assembled fibrous network. At pH 6, scattering confirms a hierarchical fibrous network with individual fibres (radius ~38 Å). At pH 8, the low-q signal ($q^{-1.6}$) indicates an open, branched 3D network with subtly thinner individual fibres (radius ~27 Å). In contrast, at pH 4, the low-q power-law ($q^{-3}$) and a substantially larger fitted radius (~424 Å) suggest the formation of large, loosely packed, disordered clusters instead of defined fibres. Yellow arrows indicate the characteristic peaks. **c & d.** TEM and AFM images of the OP-gel fibre structures at pH 4, 6, and 8. **e**. Confocal fluorescence images showing the adhesion of DiI-labeled BV-2 cells on OP-gel films under different pH conditions (pH 4, 6, and 8) after 6 hours

of incubation. **f**. Representative image demonstrating the adhesion of the OP gel to brain tissue. **g**. The adhesive properties of OP-gel films under different pH conditions (pH 4, 6, and 8) were quantified in liquid environment using AFM-based force spectroscopy. The results revealed that the OP gel exhibits stronger adhesive properties under the physiological pH conditions. **h & i**. Diameter analysis of the OP-gel fibres based on TEM and AFM images. **j**. Cell adhesion analysis based on the confocal fluorescence images showed in **e**. The result revealed the same trend as those obtained from the AFM force spectroscopy measurements. ns at $p > 0.05$; ****$p < 0.0001$ by one-way ANOVA.

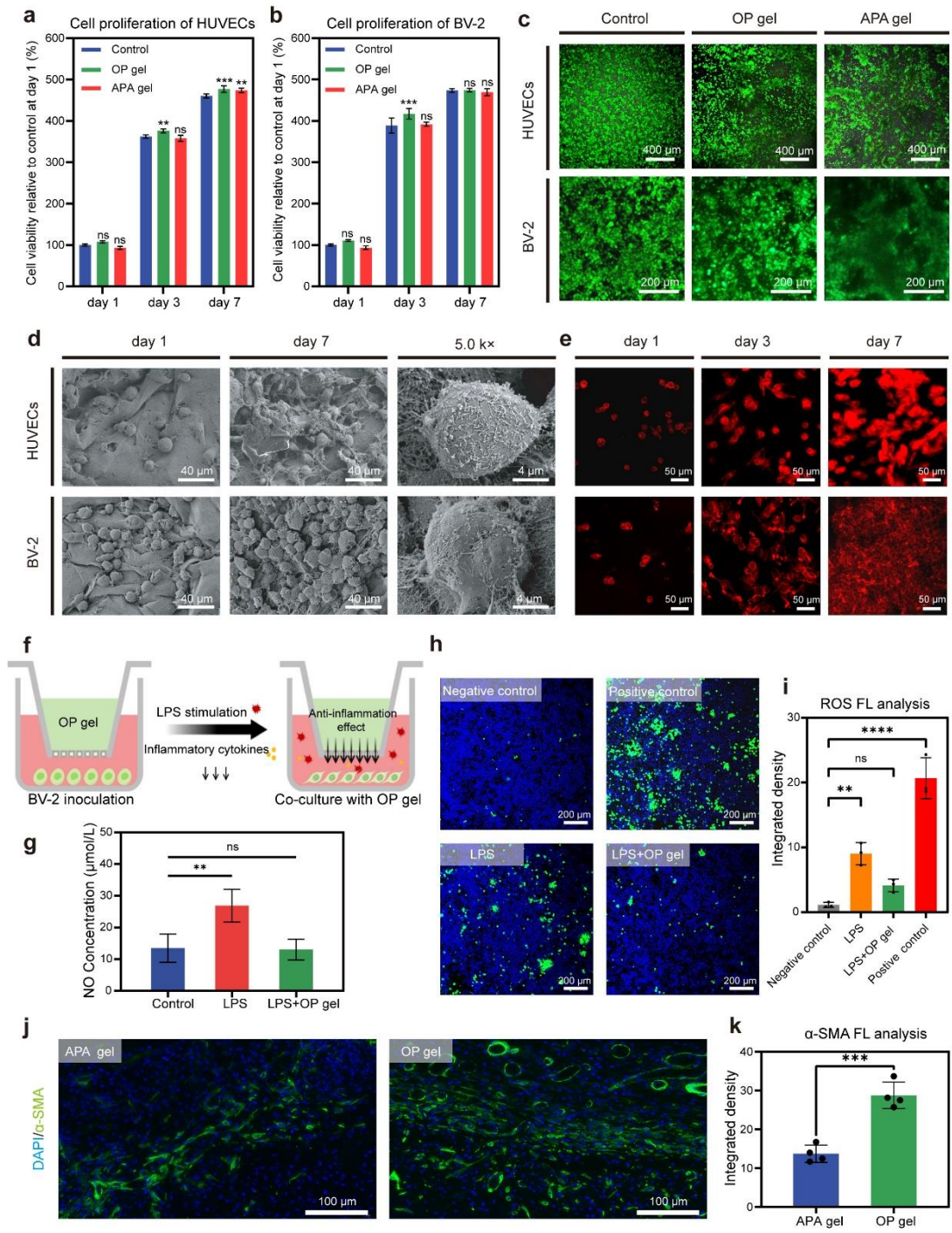

**Fig. 4 | Biological validation of the OP gel. a** & **b**. Cell viability of human umbilical vein endothelial cells (HUVECs) and BV-2 cells co-cultured with materials assessed by CCK-8 assay on day 1, 3, and 7 (absorbance at 450 nm). ns at $p > 0.05$; **$p = 0.0027$; ***$p < 0.001$ by Two-way ANOVA. **c**. Live (green)/dead (red) assay confirmed the proliferation of HUVECs and BV-2 3D co-cultured on the materials at day 7. **d**. SEM images confirmed the proliferation and 3D growth of HUVECs and BV-2 cells on the materials at day 1 and 7. **e**. HUVECs and BV-2 cells after 3D co-culture on the materials were stained with phalloidin (F-actin, red) at day 1, 3, and 7. **f**. Illustration of the lipopolysaccharide (LPS)-induced BV-2 neural inflammation model and co-culture with OP gel using a Transwell assay. **g**. Quantification of NO release revealed that the OP gel significantly

inhibited LPS-induced NO production. ns at p > 0.05; **p = 0.0022 by One-way ANOVA. **h**. Confocal images of reactive oxygen species (ROS) staining in OP-gel co-culture and control groups (green: ROS$^+$, blue: nucleus). ns at p > 0.05; **p = 0.0022; ****p < 0.0001 by Two-way ANOVA. **i**. Quantification of the ROS$^+$ fluorescence intensity revealing that the OP gel effectively reduces LPS-induced intracellular ROS generation. **j**. Immunofluorescence staining of α-SMA and DAPI in subcutaneous tissues 2 weeks post-implantation of OP gel and APA gel (green: α-SMA$^+$ microvessels, blue: nucleus). **k**. Quantification of the α-SMA$^+$ fluorescence intensity revealing that OP gel exhibits more pronounced vascularization. ***p = 0.0003 by Unpaired t test.

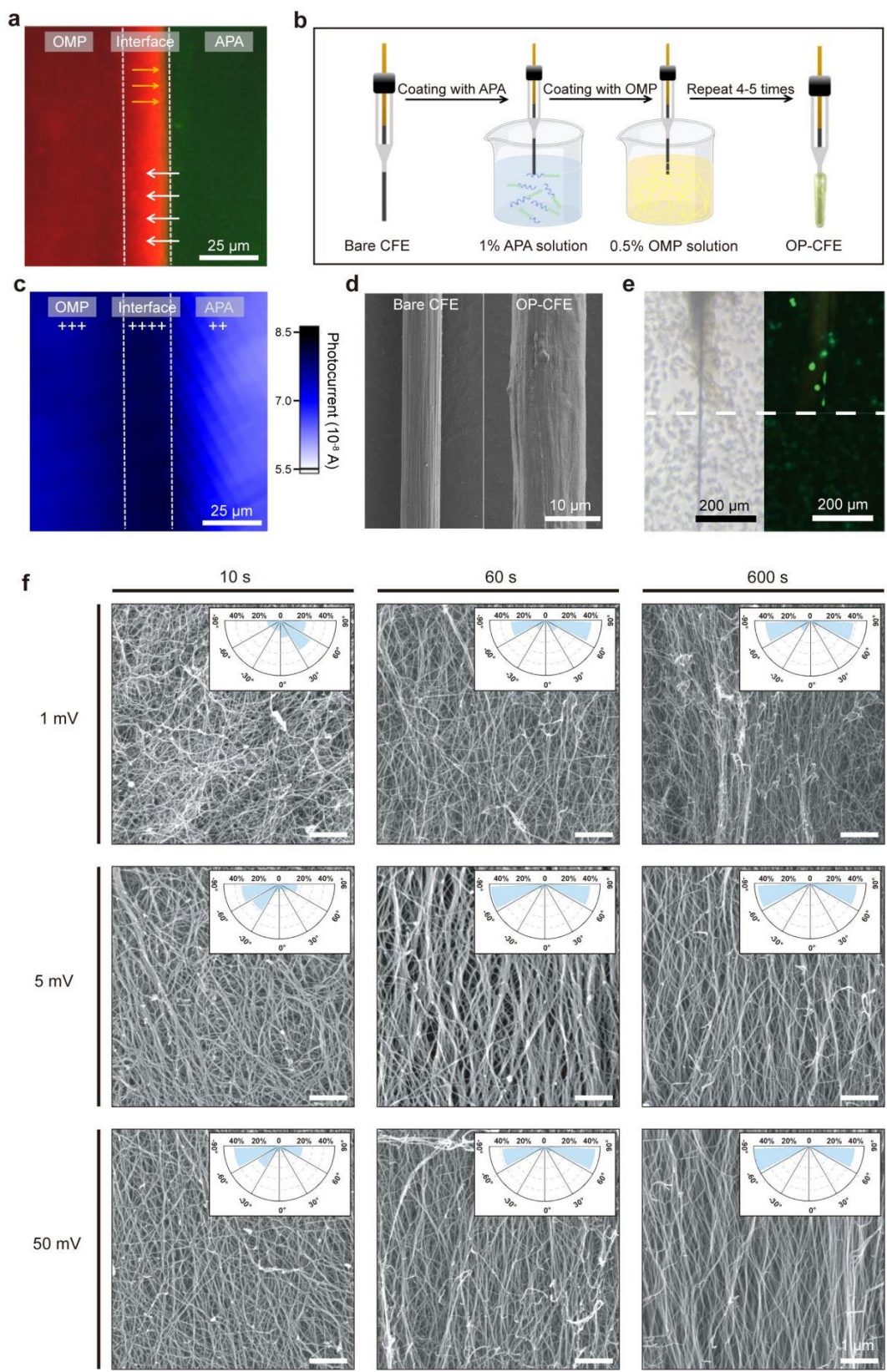

**Fig. 5 | Fabrication and interfacial assembly *in situ* of OP-CFE. a**. The confocal image of OMP-APA co-assembly interfacial (red: OMP, green: APA): LLPS-mediated interfacial layer formation (white arrows) and diffusion-reaction-controlled directional gelation (yellow arrows). **b**. Illustration

of the *in-situ* fabrication of OP-CFE *via* a facile dip-coating process. **c**. Spatial photocurrent mapping image of OMP-APA co-assembly interfacial revealing that co-assembly interfacial exhibited a higher photocurrent response than regions composed of either component alone. **d**. SEM images of the tips of bare CFE and OP-CFE. **e**. Optical microscopy image and confocal image (live/dead staining) of carbon fibres partially coated with OP gel which were co-cultured with BV-2 cells for 6 h, revealing that OP-gel coating promotes early cellular adhesion. **f**. SEM images of the fibrillar morphology of OP gel subjected to electrical stimulation at varying intensities and durations. Fibre orientation analysis revealed that the OP gel undergoes an electrically triggered disorder-to-order transition.

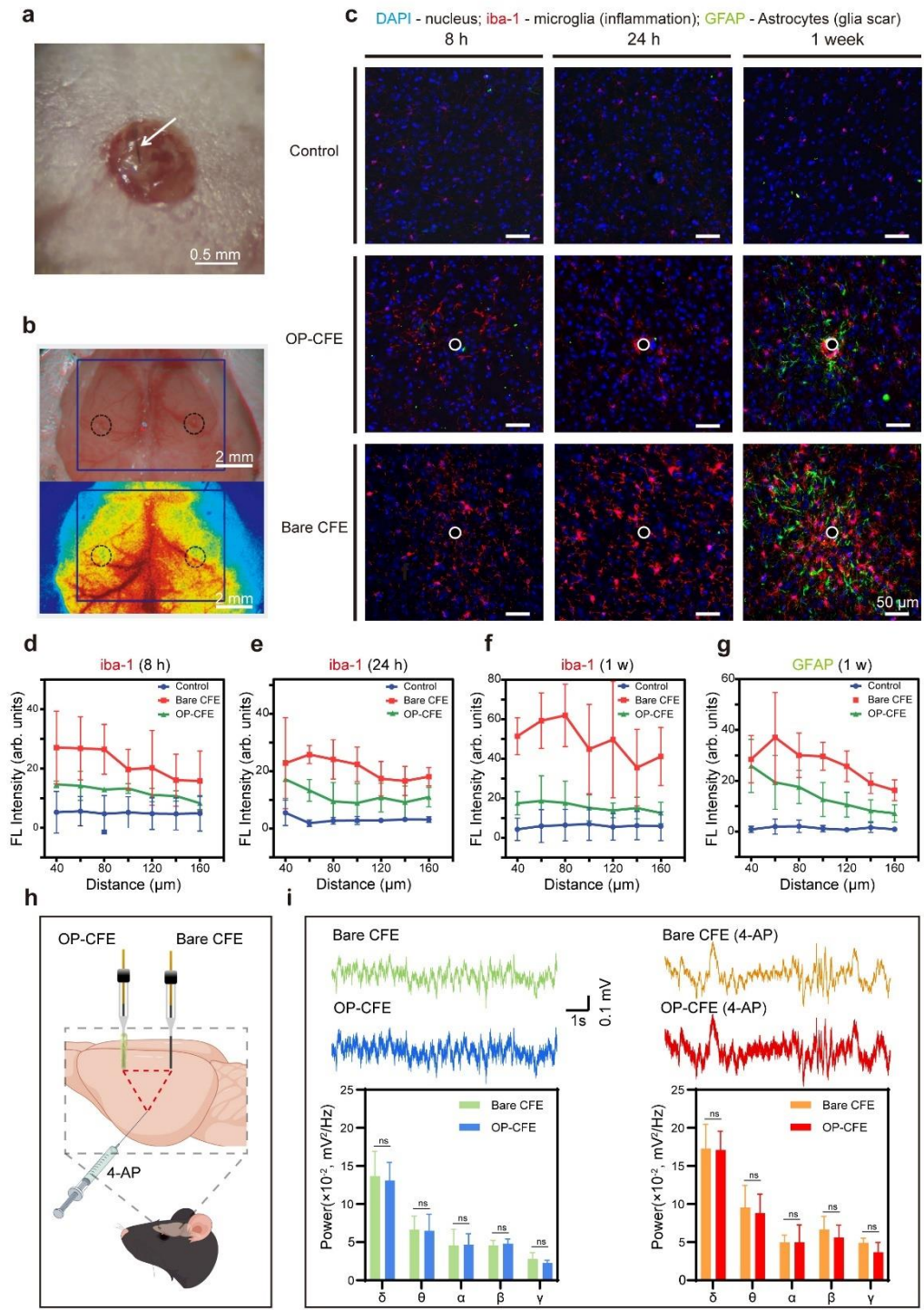

**Fig. 6 | Tissue integration and electrical performance of OP-CFE. a.** Representative image showing the precise insertion of an OP-CFE into the mouse primary motor cortex (M1). b. Cerebral blood flow (CBF) map acquired by laser speckle contrast imaging (LSCI) 24 hours after implantation of a bare CFE and an OP-CFE. The left circle indicates the site implanted with the OP-CFE, and the right circle indicates the site implanted with the bare CFE. c. Histological comparison of brain tissues from untreated mice and mice after 8 h, 24 h and 1 week implantation of bare CFE or OP-CFE in M1 cortex (n = 3, for each group). Tissues are labeled for astrocytes (green), microglia (red) and nucleus (blue). **d-g**. Fluorescence intensities of microglia (**d-f**) and astrocytes (**g**) in brain

tissues from untreated mice (blue line) and mice after implantation of bare CFE (red line) and OP-CFE (green line) (n = 3, for each group). The distance was measured from the image center for the control group and from the implantation center for the bare CFE and OP-CFE groups. **h**. Illustration of the *in vivo* electrophysiological recording and pharmacological stimulation experiment (n = 6). Positions of the two electrode implants (bare CFE, OP-CFE) and the 4-aminopyridine (4-AP) injection site are spaced 1 mm apart (center-to-center) over the M1 cortex. **i**. Local field potentials (LFPs) recorded from implanted bare CFE and OP-CFE before and after 4-AP injection (top). Quantification by power-spectral density analysis (bottom). ns at $p > 0.05$ by Two-way ANOVA.